\documentclass[twocolumn,trackchanges]{aastex63}

\received{MM dd, 2020}
\revised{MM dd, 2020}
\accepted{MM dd, 2020}
\submitjournal{AJ}

\shorttitle{Searching for active low-mass stars in CMa region}
\shortauthors{Gregorio-Hetem et al.}

\begin{document}

\title{Searching for active low-mass stars in CMa star-forming region:
multi-band photometry with T80S}

\correspondingauthor{Jane Gregorio-Hetem}
\email{gregorio-hetem@usp.br}

\author[0000-0002-3952-7067]{J. Gregorio-Hetem}
\affiliation{Universidade de S\~{a}o Paulo, IAG,
Rua do Mat\~{a}o 1226, 05508-090
S\~{a}o Paulo, Brazil }

\author[0000-0002-0284-0578]{F. Navarete}
\affiliation{Universidade de S\~{a}o Paulo, IAG,
Rua do Mat\~{a}o 1226, 05508-090
S\~{a}o Paulo, Brazil }

\author[0000-0002-1049-4812]{A. Hetem}
\affiliation{UFABC Federal University of ABC,
Av. dos Estados, 5001, 09210-580
Santo Andr\'{e}, SP, Brazil }

\author[0000-0001-8789-6230]{T. Santos-Silva}
\affiliation{Universidade de S\~{a}o Paulo, IAG,
Rua do Mat\~{a}o 1226, 05508-090
S\~{a}o Paulo, Brazil }

\author[0000-0003-2271-9297]{P.A.B. Galli}
\affiliation{Laboratoire d’Astrophysique de Bordeaux,
Univ. Bordeaux, CNRS, B18N,
All\'ee Geoffroy Saint-Hilaire, 33615 Pessac, France}

\author{B. Fernandes}
\affiliation{Universidade de S\~{a}o Paulo, IAG,
Rua do Mat\~{a}o 1226, 05508-090
S\~{a}o Paulo, Brazil }

\author[0000-0002-4304-9846]{ T. Montmerle}
\affiliation{Institut d'Astrophysique de Paris, France}

\author[0000-0002-1517-0710]{ V. Jatenco-Pereira}
\affiliation{Universidade de S\~{a}o Paulo, IAG,
Rua do Mat\~{a}o 1226, 05508-090
S\~{a}o Paulo, Brazil }

\author[0000-0001-5740-2914]{M. Borges Fernandes}
\affiliation{Observat\'orio Nacional, 
Rua General Jos\'e Cristino 77, 20921-400,
Rio de Janeiro, Brazil }

\author[0000-0002-0537-4146]{H. D. Perottoni}
\affiliation{Universidade de S\~{a}o Paulo, IAG,
Rua do Mat\~{a}o 1226, 05508-090
S\~{a}o Paulo, Brazil }

\author[0000-0002-4064-7234]{W. Schoenell}
\affiliation{GMTO Corporation 465 N. Halstead Street, Suite 250 Pasadena, CA 91107}

\author[0000-0002-8254-2959]{T. Ribeiro}
\affiliation{Rubin Observatory Project Office, 950 N. Cherry Ave., Tucson, AZ 85719, USA (AURA Staff)}

\author[0000-0002-2484-7551]{A. Kanaan}
\affiliation{Departamento de F\'isica,
Universidade Federal de Santa Catarina,
Florian\'opolis, SC, 88040-900, Brazil}

\begin{abstract}
An exotic environment surrounds the young stellar groups associated with the Canis Major (CMa)  OB1/R1 region, which probably was formed under feedback from at least three supernova events having occurred a few million years ago.
We use astrometric data from the \textit{Gaia}-DR2  to confirm the membership of the stars in CMa R1,
based on proper motion and parallax, which revealed 514 new members and candidates. 
The mean age of 5 Myr estimated from the color-magnitude diagram characterizes the sources as likely pre-main sequence candidates.
In total, a sample of  694 stars detected with  the {\it T80}-South telescope was analyzed according to different color-color diagrams, 
which were compared with theoretical colors from evolutionary models, aiming to reveal the objects that exhibit color excess due to accretion processes. 
Accretion and magnetic activity were also explored on the basis of empirical flux-flux relation, such as F$_{660}$ and F$_{861}$ that are related to 
H$\alpha$ and Ca II triplet emission, respectively. 
 A low fraction ($\sim$ 3 percent) of the sample have H$\alpha$ excess and other colors expected for stars exhibiting chromospheric activity. 
The number of Class I and Class II objects, identified by the infrared ({\it WISE}) colors, indicates a disk fraction of $\sim$ 6 percent, which is
lower than the expected for stellar clusters with similar age.
A such large sample of objects associated with CMa R1 without evidences of circumstellar accretion 
can be interpreted as a lack of disk-bearing stars, unusual for young star-forming regions. However, 
this may be explained as the result of supernova events.
\end{abstract}



\section{Introduction} \label{sec:intro}

Depicting different scenarios of star formation can bring us important clues for answering
some of the major problems in astrophysics, mainly related with the early evolution of stars and protoplanetary
disks, with implications on planet formation as well as on the structure and evolution of the Galaxy.

Despite the impressive progress that star formation study has achieved thanks to the
advances on multi-band observational data analysis, there are still many open questions related
to the physical relation between young star clusters and their respective parental clouds.

Our long-term goal is to study the young stellar population in different Galactic regions, and
investigate the influence of the environment and feedback from massive stars (ionization, winds, supernovae) in
the formation and evolution of star clusters and circumstellar disks. This work
is focused on the star-forming region Canis Major  (CMa) R1 region, aiming to improve the census of the
young stellar population. 

The CMa OB1/R1 Association is a large ($\sim$ 100 deg$^2$) complex of
molecular clouds, emission and reflection nebulae showing evidences of star formation induced by
supernova explosions. In a search for emission-line stars, \citet{Schev99} found about two hundred B stars
associated with the reflection nebulae that constitute the CMa R1 region, located at $d \sim 1$\,kpc \citep[for a review, see][]{GH08}.

CMa R1 has a mixing of young objects with ages ranging from 1 to 10
Myr, distributed on a large scale region \citep{GH09}.
Based on X-ray observations in the direction of the Sh\,2-296 nebula, \citet{Santos18} reported more than 300 low-mass young stars candidates. As noted by
\citet{Fernandes15}, among the T Tauri stars found by them around Sh~2-296, less than 10 percent
show evidence of circumstellar disks. Such low fraction
of disk-bearing stars is quite rare when compared with other star-forming regions
 \citep{Haisch01,Hernandez08,Fedele10,Cloutier14,Briceno19}, suggesting that some external
factor accelerated the disk dissipation. More recently, \citet{Fernandes19} identified
three runaway stars associated with bow-shock structures in CMa OB1/R1 region.
Those authors found that the runaway stars have
 likely been ejected from three successive SN explosions ($\sim 6$, $\sim 2$ and $\sim 1$ Myr ago), which
 might have originated the arc-shaped structure named ``CMa shell" by them.

In this work we explore the importance of large-scale, multi-wavelength surveys to characterize the 
 pre-main sequence stars associated with CMa R1 in order to investigate aspects such  as disk evolution.  
 The goal is to analyze  a large reliable sample of  members, providing a new optical characterization 
complemented with the classification based on infrared-excess. By these means, the fractions of accreting stars and 
disk-bearing stars can be inferred and compared with 
other stellar clusters with similar age.

The paper is presented as follows. 
In Sect. \ref{sec:obs} we describe the data that we obtained from previous works, and from 
observations developed by us with the {\it T80}-South telescope. We also use proper motion and parallax
from the {\it Gaia}-DR2 catalog to select the sample containing the probable members associated with CMa R1.
The criteria for sample selection and classification are presented in Sect. \ref{sec:select}, while
the methods based on diagnosis of accretion and magnetic activity, which were adopted to analyze the 
sample, are described in Sect. \ref{sec:sec4}. Finally, in Sect. \ref{sec:sec5} we summarize and discuss the main results.

\begin{figure*}
\begin{center}
\includegraphics[angle=0,width=14cm]{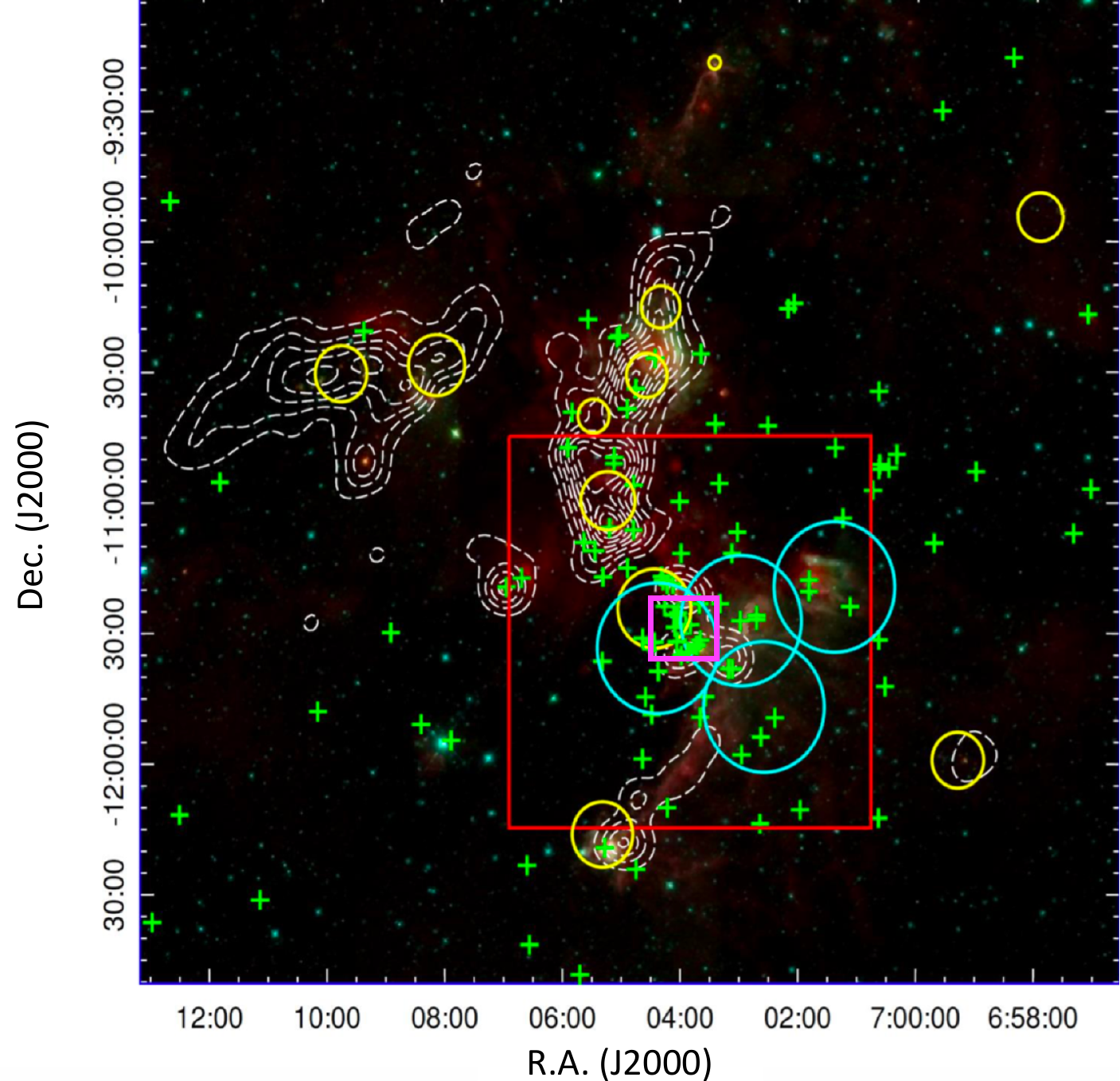}
\caption{
Infrared ({\it WISE, 22 $\mu$m}) image of CMa R1 region overlapped with $^{13}$CO map (dashed contours) from Osaka group \citep[see][]{Onishi13}, showing the position of B-type stars \citep[green crosses,][]{Schev99}, and groups of 
Class I and Class II stars \citep[yellow small circles,][]{Fischer16}. The fields covered by our observational campaigns are also indicated: {\it T80S} (red large box), 
{\it XMM-Newton} (cyan  circles),  and GMOS (pink small box).}
\end{center}
\label{fig:wise}
\end{figure*}

\begin{figure*}
\begin{center}
\includegraphics[angle=0,width=7.5cm]{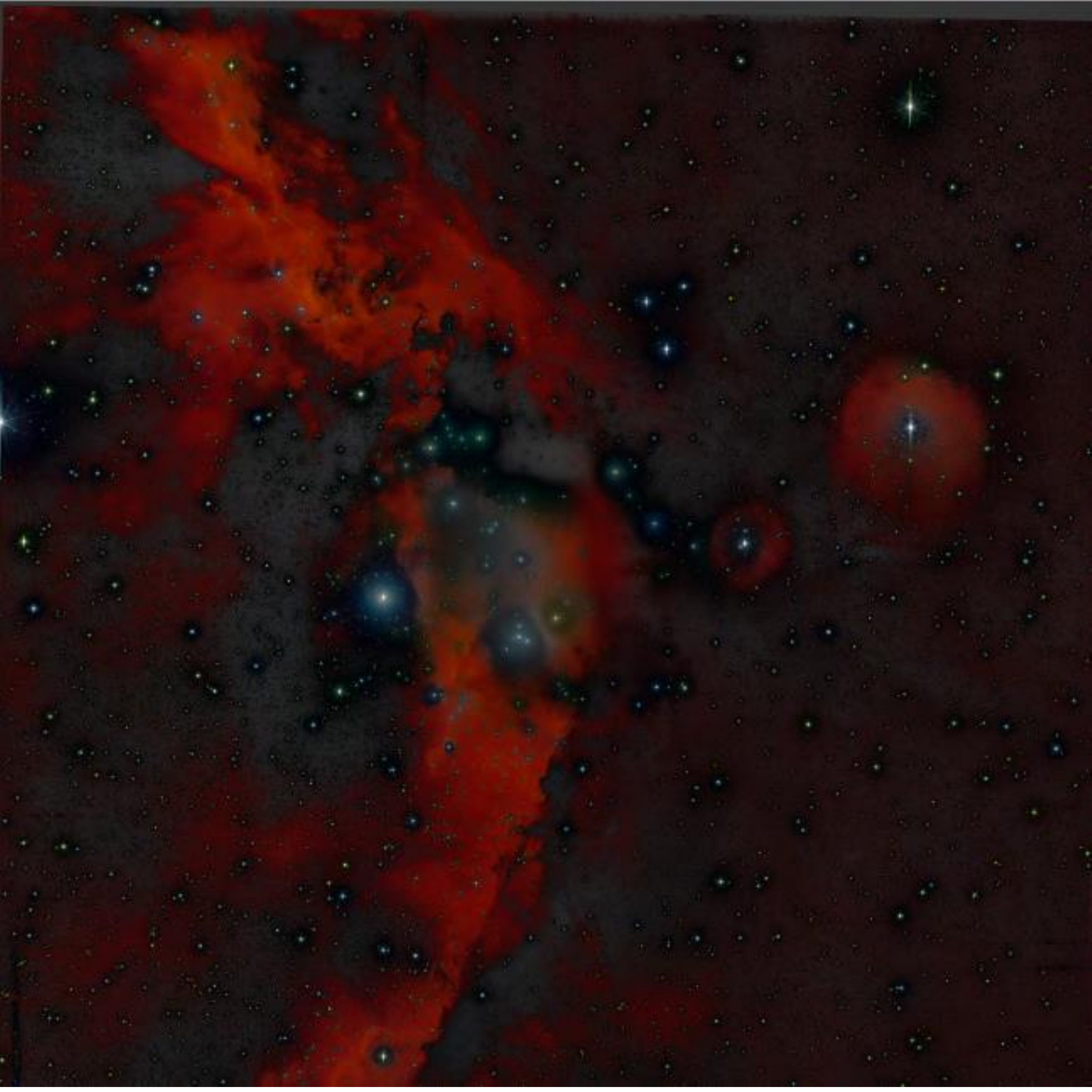}
\includegraphics[angle=0, width=8.5cm]{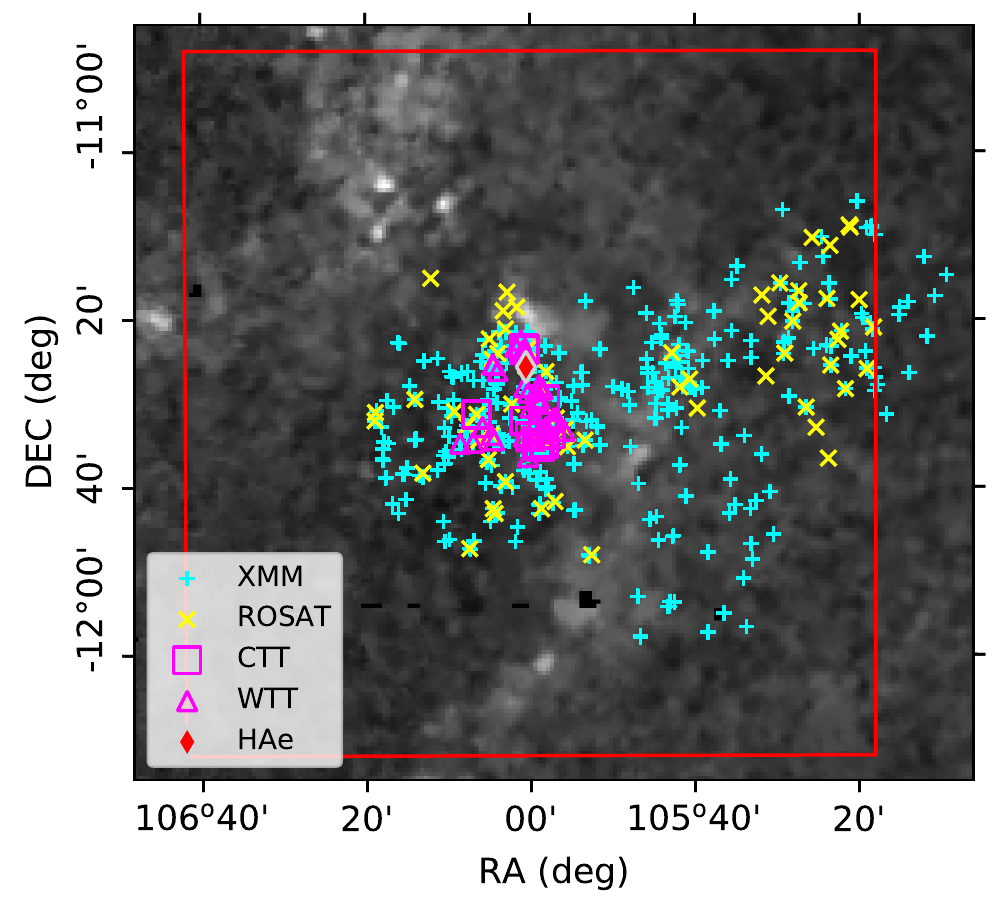}
\caption{{\it Left panel}: Combined image of filters J0660 (H$\alpha$, red), {\it r'} (green), and {\it g'}  (blue) obtained with {\it T80S} observations.
The field has 1\fdg4 each side. North is up and East is to the left. 
{\it Right panel:}   Visual extinction (A$_V$) map of CMa R1 overlapped by the T80S FoV (red square),  and the position of X-ray sources from
\citet[][{\it XMM-Newton}]{Santos18} and \citet[][{\it ROSAT}]{GH09}. Pink symbols show the previously known PMS stars \citep[][{\it GMOS}]{Fernandes15}. 
Inverse scale of colors is used for the map, where dense regions appear in white and light grey  indicates  low level of extinction (A$_V<$1 mag).}
\end{center}
\label{fig:fig2}
\end{figure*}


\section{Observations}
\label{sec:obs}

\subsection{Data from previous works}

To better understand the star formation scenario and the early evolutionary stages of the stars associated with
 CMa OB1/R1  we previously analyzed the X-ray sources detected by {\it ROSAT} and {\it XMM-Newton} satellites in the direction of the nebula Sh~2-296 \citep{GH09,Santos18}. 
The characterization of the X-ray sources was based on {\it 2MASS} data of their near-infrared counterparts for which
mass and age were estimated. 
This analysis indicates they are probably pre-main sequence stars, but their nature needs to be
confirmed by means of complementary optical data. 

For a small part (12 percent) of the area where the X-ray sources are found,  \citet{Fernandes15} obtained medium-resolution optical spectra with GMOS (multi-object spectroscopy) at the Gemini telescopes. 
The GMOS data were used to obtain the spectral classification of the candidates
 as a diagnosis of their youth nature based on the H$\alpha$ and lithium lines. The near- to mid-infrared data
 extracted from {\it 2MASS} and {\it All-WISE} catalogues were used to search for evidences of circumstellar structure, which had indicated
a low fraction of disk-bearing stars as reported by \citet{Fernandes15}. A similar result was found by \citet{Fischer16} who adopted the criteria from \citet{Koenig14} for distinguishing Class I, II and III objects that respectively are embedded, disk-bearing, and disk-lacking stars.
\citet{Fischer16} found a few groups of Class I or Class II stars over the entire area of CMa OB1/R1.
These groups are shown in Fig. \ref{fig:wise}, together with the distribution of dust (WISE image) and molecular clouds (CO map) of the region.

Figure \ref{fig:wise} also shows the regions covered by different surveys performed by us to study the CMa R1 region: optical spectroscopy with Gemini telescopes \citep{Fernandes15}, {\it XMM-Newton} \citep{Santos18}, and optical photometry obtained with {\it T80}-South telescope (see Sect. \ref{sec:t80}). In spite of the fact that intermediate-mass stars are well
known in this region \citep{Schev99,GH09}, the sample of fainter, low-mass stars is still incomplete.

According to the literature, in the area covered by the {\it T80}-South observations (red square in Fig. \ref{fig:wise}),  there are 430 stars that are likely associated with the CMa R1
region. Among those previously known objects, 60 corresponds to B- or A-type stars \citep{Claria74,Schev99} and 370  are young  low-mass stars candidates selected from {\it XMM} data \citep{Santos18}.
However,  64 percent of them  does not fulfill  the parallax and kinematic criteria used to confirm their association with CMa R1 (see further discussion in  Sect. \ref{sec:pm}).

Therefore, the optical observations from {\it T80}-South  combined with {\it Gaia} data analyzed in this work aim both purposes:  $(i)$ complementing the characterization of 
the candidates found by {\it XMM}  around Sh~2-296, which is necessary to confirm if they are Pre-Main Sequence
 (PMS)  stars, and $(ii)$ seeking the detection of 
young star candidates in a larger area  of the CMa R1 region. Figure \ref{fig:fig2} (right panel) provides a comparison of the areas covered by our observations showing the distribution of candidates  that  are X-ray sources ($\sim 0\fdg5 \times 1\fdg3$),  and the confirmed PMS stars, which were observed by 
GMOS  ($\sim 0\fdg23 \times 0\fdg33$). 
This illustrates the gain of {\it T80}-South observations to identify PMS member candidates across this OB association in a much larger area than the previous follow-up spectroscopic study of \citet{Fernandes15}.

\subsection{Multi-band photometry with {\it T80S}}
 \label{sec:t80}

Optical imaging in a wide (1\fdg4 $\times$ 1\fdg4) field-of-view (FoV), 
covering the CMa R1 region was performed in 2017, 17 January,
with the 0.8-meter telescope {\it T80}-South (hereafter {\it T80S}) located at Cerro Tololo (Chile).
{\it T80S} is dedicated to
performing the ``Southern Photometric Local Universe Survey" (S-PLUS) \citep[for more details see][]{Claudia19}.
S-PLUS uses five broadband SDSS filters (\textit{u', g', r', i', z'}) and seven narrow-band filters centered on features particularly interesting for our study, like
Ca II H+K ($\lambda$\,=\,3968.5\AA, \,3933.7\AA), H$\alpha$ ($\lambda$\,=\,6562.8\AA), and Ca II infrared triplet
(IRT, $\lambda$\,=\,8498\AA, \,8542\AA, \,8662\AA),  among others. A set of three images was obtained for each of the 12 filters. 
Table \ref{tab:log} presents details of data acquisition in the CMa R1 region.

 Figure \ref{fig:fig2} (left panel) shows a false-color composite RGB image combining three filters, enhancing the nebular (H$\alpha$) emission 
associated with Sh~2-296. For comparison with the dust distribution in the cloud, the right panel of Fig. \ref{fig:fig2} displays the visual  extinction map \citep{GH08} towards the area covered by {\it T80S}. The position of the X-ray candidates is also shown to indicate the area covered by  {\it ROSAT} and {\it XMM} observations 
\citep{GH09,Santos18}. Different symbols are used to show were the previously known PMS stars are located. 

The data\footnote{Data reduction is made using the J-PLUS pipeline \citep{cenarro}} extraction and calibrations procedures were performed by the S-PLUS team, which developed a pipeline
based on the {\it SExtractor}  software \citep{Bertin96, Bertin10} to construct the photometric catalogues \citep{Claudia19}. The S-PLUS pipeline
provides a first estimate of the flux  corresponding to the zero-point of magnitude, to be used in the conversion of instrumental magnitudes.

By adopting the mean values of zero-point
(ZP$_o$) magnitudes estimated by the S-PLUS team{\footnote{ ZP$_o$ calculated by L. San Pedro (private communication).}} as a first guess, we obtained a preliminary catalogue of about
73700 sources extracted from the {\it T80S} images. In Sect. \ref{sec:calib} we present details on the photometric calibration and selection of the sub-sample that is analyzed in this work.
The astrometry of the detected sources was checked by comparing their position with the
known emission-line stars from \citet{Schev99}.

\begin{figure*}
\includegraphics[angle=0,width=18cm]{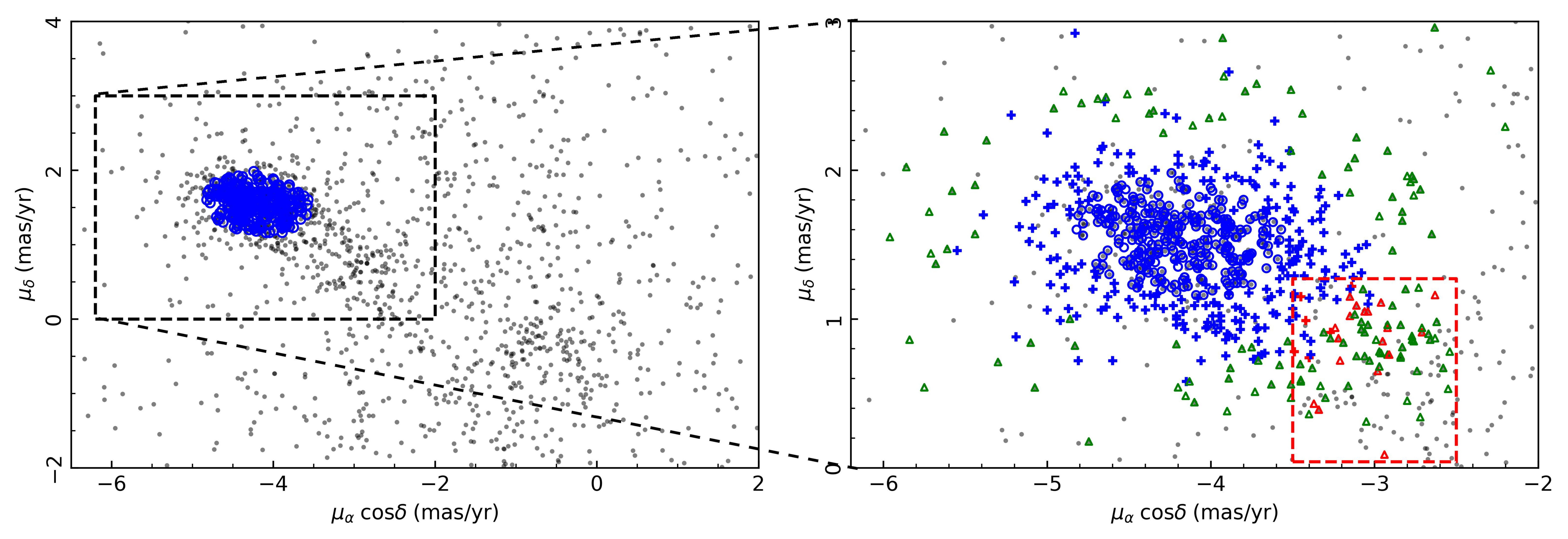}
\caption{{\it Left:} Sources selected from \textit{Gaia-DR2} (gray dots) for which we estimate membership 
probabilities based on proper motion. Blue circles indicate very-likely members with P $\geq$ 90 percent. 
{\it Right:} a zoom of the main clustering of objects observed by both GAIA and {\it T80S} surveys. The  membership 50 $\leq$ P$<$ 90 percent
defines  the probable members (blue +), while  and  candidates (green triangles) have P $<$ 50 percent. 
The sources in the apparent sub-cluster (dashed red box) may be part of a group (highlighted by red symbols) that is located to the NW direction of the region (see Fig. \ref{fig:radec}.)
}
\label{fig:pmov}
\end{figure*}

\section{Sample selection and classification}
\label{sec:select}

The main selection criterion is based  on  the second GAIA data release \citep[{\it Gaia-}DR2,][]{Gaia18}, which was used   to confirm the membership of the objects associated with the nebula, and to exclude sources detected by the {\it T80S}  that are likely field-stars in the line-of-sight of the CMa R1 region (Sect. \ref{sec:pm}).

Infrared  and optical colors  were used in order to classify the new PMS members and candidates,  as well as the previously known candidates, in
comparison with confirmed PMS objects.
Firstly we search for sources exhibiting  infrared (IR) excess based on data from the {\it WISE} survey  (Sect. \ref{sec:class}),  and secondly we estimate ages
using the {\it Gaia}-DR2 (Sect. \ref{sec:age}).

\subsection{Selection of kinematic members}
 \label{sec:pm}

The query was restricted to the sources in the
 parallax range  $\varpi$ = 0.8 - 1.25 milliarcseconds (mas) that is consistent with the distance of the 
 cloud \citep[{\it d} $\sim$ 1kpc,][]{GH08}. To avoid objects showing low quality of  the astrometric solution, 
 we applied the selection criteria of 
$\varpi / \sigma_{\varpi} > 3$ and RUWE{\footnote{Re-normalised unit weight error (see details in the technical \\
note GAIA-C3-TN-LU-LL-124-01)}} $< 1.4$ that  resulted in an initial list of 4153 objects found in the {\it T80S} FoV. 

For estimation of membership probability (P)  we adopted the method from \citet{Hetem19}
combining a Bayesian model and the Cross Entropy (CE) technique to identify cluster members based on their proper motion ($\mu_{\alpha}\cos\delta$, $\mu_{\delta}$).
Due to the high sensitivity of this method to initial conditions, we used a Genetic Algorithm
presented by \citet{HGH07} to estimate optimized initial values required to create the first generation of parameters for the CE.

The result of the CE method gives the range of proper motion
$\mu_{\alpha}\cos\delta = -4.1 \pm 0.6 ~$mas yr$^{-1}$,
$\mu_{\delta} = 1.5 \pm 0.4 ~$mas yr$^{-1}$ that was adopted as criterion to select the stars  probably associated with the CMa R1 region. 
 According to the proper motion criterion, 78 percent of the objects extracted from {\it Gaia}-DR2 probably are field stars (P$< $1 percent). The remaining list to be studied contains about 
 900 {\it Gaia} sources, 281 of them are very-likely members (P $\geq$ 90 percent), which can be seen in the 
 $\mu_{\alpha}\cos\delta$-$\mu_{\delta}$ plot presented in  Fig. \ref{fig:pmov} (left panel).

To produce the final list of objects, we performed a cross-matching of the sources selected from  {\it Gaia} with our 
preliminary catalog of objects detected with {\it T80S}.
For the  281 {\it Gaia} sources  with membership P $\geq$ 90 percent, we have found  208 cross-matching pairs in the 
{\it T80S} catalog, meaning that 73 very likely members  were missed in the source extraction
 performed by the {\it T80S} team that uses an automatic pipeline, which is based on aperture photometry. We adopted the Star Finder (SF)
 procedure \citep{Diolaiti2000} on three {\it T80S} images (filters J0430, J0515, and J0660),
 seeking for sources that are absent in our preliminary catalog.
 
\begin{figure}
\includegraphics[angle=0,width=\columnwidth]{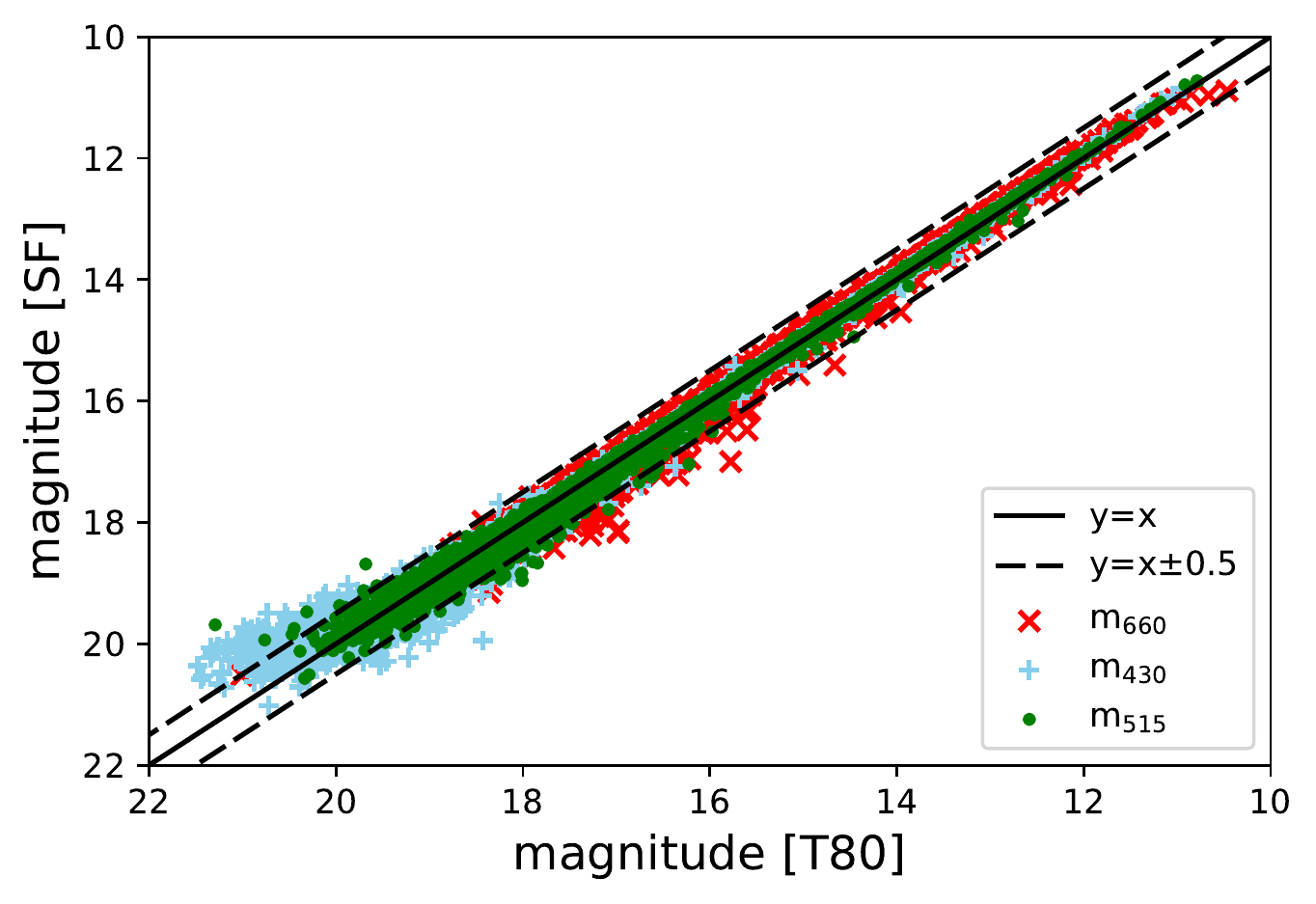}
\caption{Comparison of magnitudes obtained with filters J0660, J0515 and J0430, by using two different extraction methods:
automatic procedure performed by the {\it T80S} team and StarFinder (SF) routine adopted by us. A dispersion is found for the faintest sources  ($m_{430} >$20 mag.)}
\label{fig:sf}
\end{figure}

The cross-matching of our preliminary catalogue
and the list obtained with SF is  illustrated in Fig. \ref{fig:sf}. The magnitudes found by adopting both techniques are in
agreement for sources brighter than 20 mag at 660 nm, showing a dispersion of 0.5 mag (or more) in the case of
fainter magnitudes. This result gives us confidence in considering in our analysis the list of  sources produced by SF, 
which  were missed in the preliminary catalogue. These additional sources were included in our list 
that contains  669 selected objects: 281 members (P $\geq$ 90 percent), 242  probable members (50 $\leq$ P $<$ 90 percent),  and 146 candidates (P $<$ 50 percent).

Figure \ref{fig:pmov} (right panel) illustrates our selection criterion that is based on the above mentioned range of proper motion. 
In the plot centre there is a main clustering where the very-likely members are concentrated (inside an ellipse of 1$\sigma$ size). 
Around  the main clustering we selected the probable members and candidates, which  are contained in a 3$\sigma$ distribution that overlaps an  
apparent sub-cluster (seen to the right). 
Objects in this sub-cluster  may be part of a group located to the NW direction that was identified by Santos-Silva et al. (2020, hereafter SS20) 
based on a clustering code that uses 5-dimensional data from Gaia-DR2 to diagnose physical groups in the direction of the CMa OB1 stellar association.

Table \ref{tab:pm} summarizes the astrometric and kinematic parameters of our sample, compared with the results for two groups
found by SS20. The proper motion and parallax of those groups coincide with the range adopted here to select members and candidates:
our main clustering has parameters similar to those estimated  by SS20 for the group called CMa06,
while the  sub-cluster seen in Fig. \ref{fig:pmov} may be part of  other group called CMa05.

\begin{figure}
\includegraphics[angle=0,width=\columnwidth]{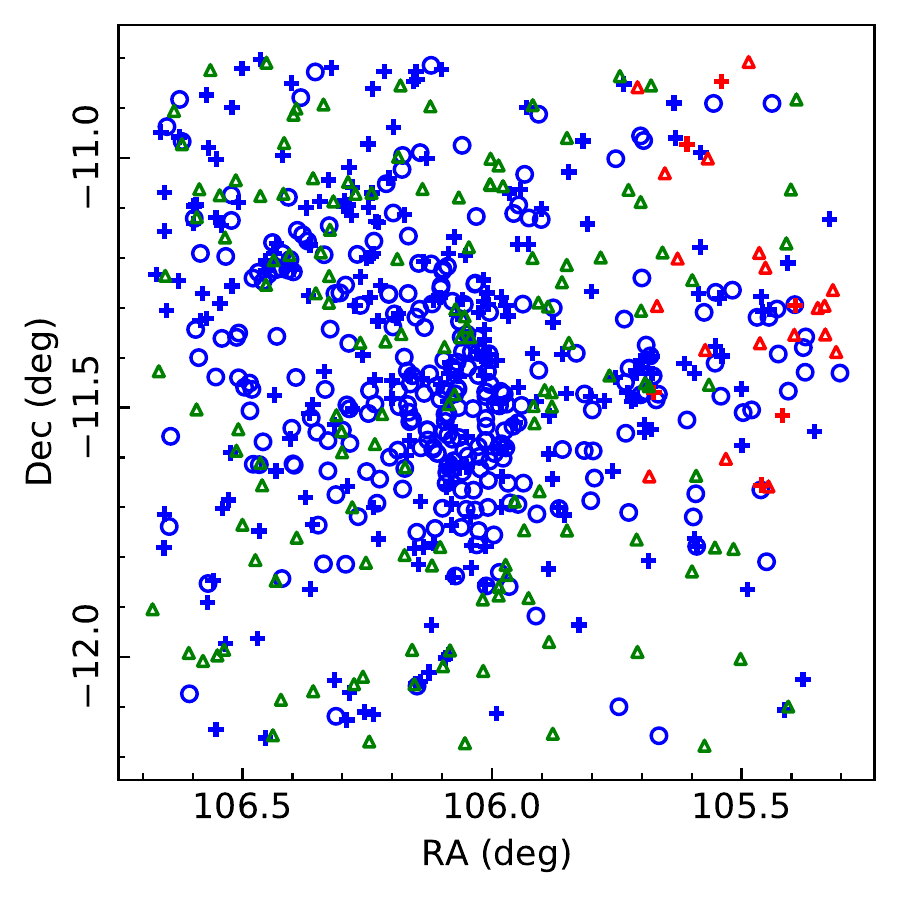}
\caption{Distribution of candidates and members in the area covered by {\it T80S} observations. The symbols are the same used in Fig. \ref{fig:pmov}.
 In the NW region we use red symbols to highlight the sources  appearing  in the sub-cluster seen in Fig. \ref{fig:pmov}. }
\label{fig:radec}
\end{figure}

A view of the distribution of the sample in the equatorial coordinates space is presented in Fig. \ref{fig:radec}.
As expected, it can be noted  a clear main concentration of sources at the center, 
tending to the E direction of the observed region,
and  roughly following the arc-shaped nebula shown in Fig. \ref{fig:fig2}. 
 According to \citet{GH09}, there is a mixing of objects with ages from 1 to 10Myr in this area that includes the sources at the NW direction.
 In Sect. \ref{sec:age} we revisit these population differences, exploring our sample of young candidate members that is larger than those previously studied.

The list of objects that were previously identified in the literature and were extracted from the {\it Gaia}-DR2 according to our selection criteria contains 
155 objects. Table \ref{tab:lista2}  gives the kinematic data for 149 of these known objects that were also identified in our {\it T80S} catalog. 
The last lines of Table \ref{tab:lista2}  present the six objects not found in the {\it T80S} catalog, 
probably due to saturation (for bright objects),  low S/N photometry in the case of faint sources, or nebular contamination.

\begin{deluxetable}{lccccc}
\label{tab:pm}
\tablecaption{Parameters related to the {\it Gaia}-DR2 data analysis}
\tablewidth{700pt}
\tabletypesize{\scriptsize}
\tablehead{
\colhead{ID} & \colhead{RA}   & \colhead{Dec}    &  \colhead{$\varpi$}           &\colhead{$\mu_{\alpha}\cos\delta$} &\colhead{ $\mu_{\delta}$} \\
\colhead{}           & \colhead{J2000} & \colhead{J2000} & \colhead{mas}                  & \colhead{$mas yr^{-1}$}               & \colhead{$mas yr^{-1}$}
}
\startdata
{\it T80S}$^{a}_{total}$ &  106\fdg0$\pm$0.7  &  -11\fdg5$\pm$0.7 &  0.8 - 1.25 &  -4.1$\pm$0.6 & 1.5$\pm$0.4 \\ 
{\it T80S}$^{a,b}_{sub}$ &  105\fdg5$^{+0.2}_{-0.2}$    &  -11\fdg3$^{+0.5}_{-0.3}$ & 0.9$\pm$0.1 &   -3.1$\pm$0.2 & 0.9$\pm$0.3 \\ 
\hline
CMa06$^c$ &  106\fdg2$\pm$0.2  &  -11\fdg4$\pm$0.3 &  0.85$\pm$0.09 &  -4.2$\pm$0.4 & 1.5$\pm$0.2 \\ 
CMa05$^c$ &  105\fdg3$\pm$0.2  &  -11\fdg2$\pm$0.2 &  0.76$\pm$0.09 &  -3.0$\pm$0.1 & 0.8$\pm$0.1 \\ 
\enddata
\tablecomments{
(a) This work; (b) sub-structure seen in Fig. \ref{fig:pmov}; (c) Results from other work (SS20) that identified two clusters coinciding with our sample.
}
\end{deluxetable}


For comparison purposes, we also analyzed 37 confirmed PMS stars, which were classified by \citet{Fernandes15}  according to H$\alpha$  and Li lines detected in {\it GMOS} spectra. 
However,  only 6 of them are listed  in Table  \ref{tab:lista2}, while  the other sources do not fulfill our selection criteria, and/or have
poor photometry (mentioned above). 
To take in  account the other 31 known PMS stars that were missing, it was necessary to relax the kinematic criteria in the Gaia data query, 
and also to adopt a lower threshold limit to  extract their  {\it T80S}  fluxes. Table \ref{tab:lista3} gives the list of these objects: 7 Classical T Tauri stars (CTT), 
23 weak-line T Tauri stars (WTT), and one Herbig Ae star (HAe).

In total, our final sample contains  694 objects, which are analyzed in this work. Besides the  180 stars previously identified (149+31, above mentioned), there are 514 objects that we suggest to be new young stars associated with CMa R1.
Among the new stars identified in this work,
395 are probable members (P$\geq$ 50 percent) for which we present the {\it Gaia}-DR2 data in Table \ref {tab:lista4},
while 119 are candidates (P$<$ 50 percent) presented in Table \ref {tab:lista5}.


\subsection{Infrared excess}
 \label{sec:class}

Part of our sample was previously studied in order to search for disk-bearing stars among the  candidates revealed by {\it XMM-Newton} data from \citet{Santos18},  focusing on 205 X-ray sources exhibiting a single IR counterpart \citep{GH16}.
The IR colors  were compared to the criteria  suggested by \citet{Koenig14} to distinguish different PMS classes  using 
 {\it WISE} photometric bands \citep{Wright10}, according with the distribution of  stars studied in Taurus by \citet{Rebull10}. 
Only a small fraction (16/205) of our X-ray sources  were considered disk-bearing candidates, according to their position in the  Class II expected locus. Most (92\%) of the IR counterparts of the {\it XMM} sources in CMa have {\it WISE} colors of diskless  (Class III) objects.

\begin{figure}
\includegraphics[angle=0,width=\columnwidth]{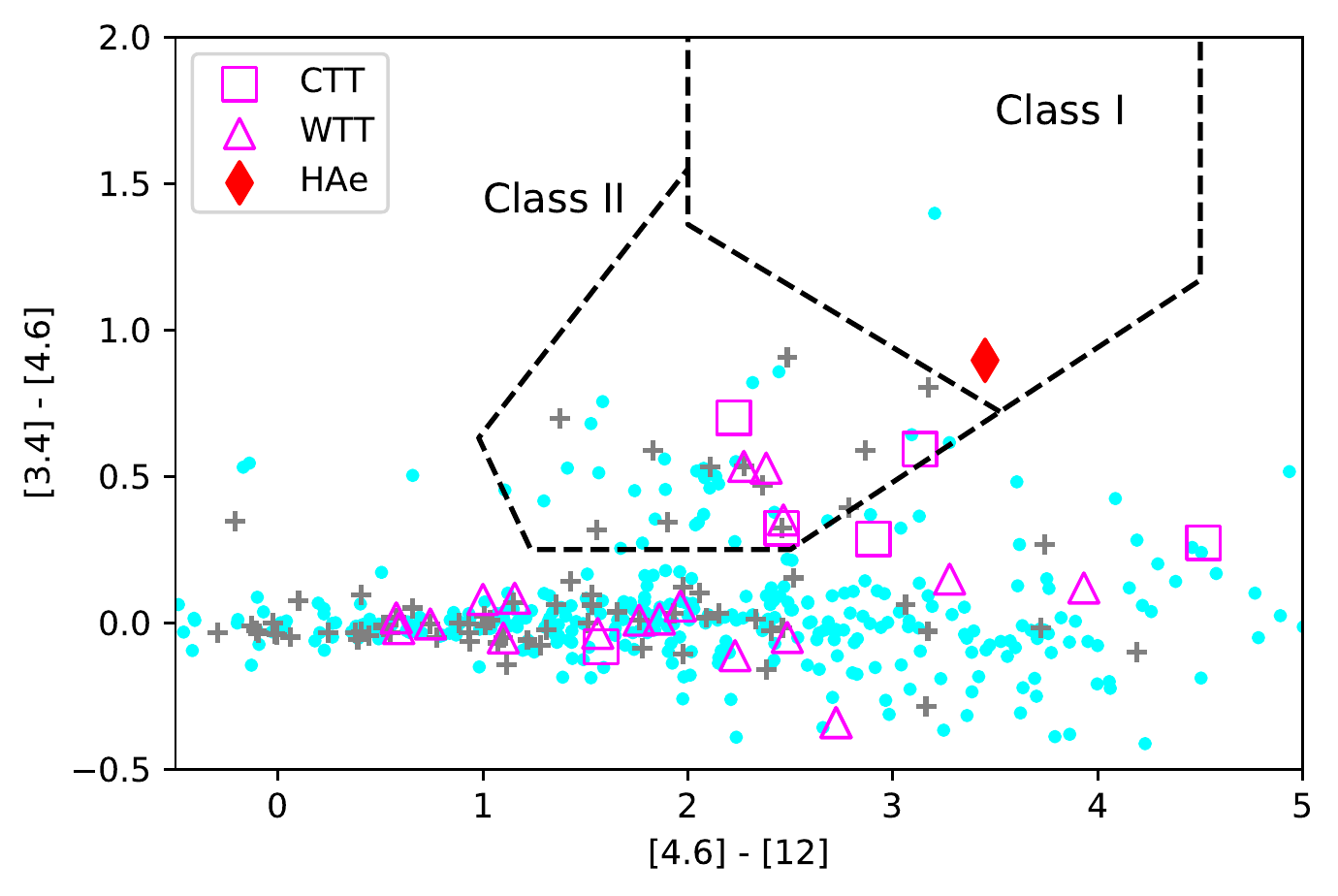}
\caption{{\it WISE} color-color diagram obtained for our sample: new members (cyan dots);  previously known candidates (grey $+$ ); and the
T Tauri  stars (pink) and one HAe star (red diamont) spectroscopically confirmed by  \citet{Fernandes15}.
Dashed lines indicate the expected locus of Classes I and II  young stars (Koenig \& Leisawitz 2014). 
The objects outside these boundaries probably are Class III (disk-lacking) stars.
}
\label{fig:wise_cor}
\end{figure}

 Here, we extend the same analysis for a larger and more reliable sample in which membership is confirmed by kinematic criteria. 
 Figure \ref{fig:wise_cor}  shows the  $[3.4]-[4.6] \times  [4.6]-[11.6]$ 
 diagram  of 79 known objects from Table \ref{tab:lista2}, and 357 new members and candidates (Tables  \ref{tab:lista4} and \ref{tab:lista5})
  identified with good IR photometric data in the  {\it All-WISE} Catalog, according with the mitigation and contamination filters 
  proposed by \citet[][see their Sect. 2.3]{Koenig14}.
 We also plot 26  previously confirmed PMS stars,  but in this case, no photometric quality filter was used. 
 Only two objects coincide with the Class I region: a new member, and the Herbig Ae star identified by \citealt{Fernandes15}).  
 The Class II locus in Fig. \ref{fig:wise_cor} is occupied by six T Tauri stars; 11 known candidates;  
 and  24 new members (Tables \ref{tab:lista4} and \ref{tab:lista5}). In summary, a total of 43 objects 
 were identified in Fig. \ref{fig:wise_cor}, from which 41 correspond to Class II and two are Class I objects. 
 These sources exhibiting IR excess are highlighted in the following figures.

 Taking into account the number of objects with available {\it WISE} photometry (463 objects), the fraction of disk-bearing stars (Classe I or II) is 
  43/463 $\sim$ 9 percent. 
  This is an upper limit, since there are other 206 objects of our entire sample that do not have {\it WISE} data. It is tempting to suggest  these remaining stars probably are Class III, assuming that a lack of data could be due to fainter IR emission. 
 If this hypothesis is correct, the fraction of disk-bearing stars  drops to 43/694 $\sim$ 6 percent.
 
\subsection{Ages}
 \label{sec:age}
\begin{figure}
\includegraphics[angle=0,width=8cm]{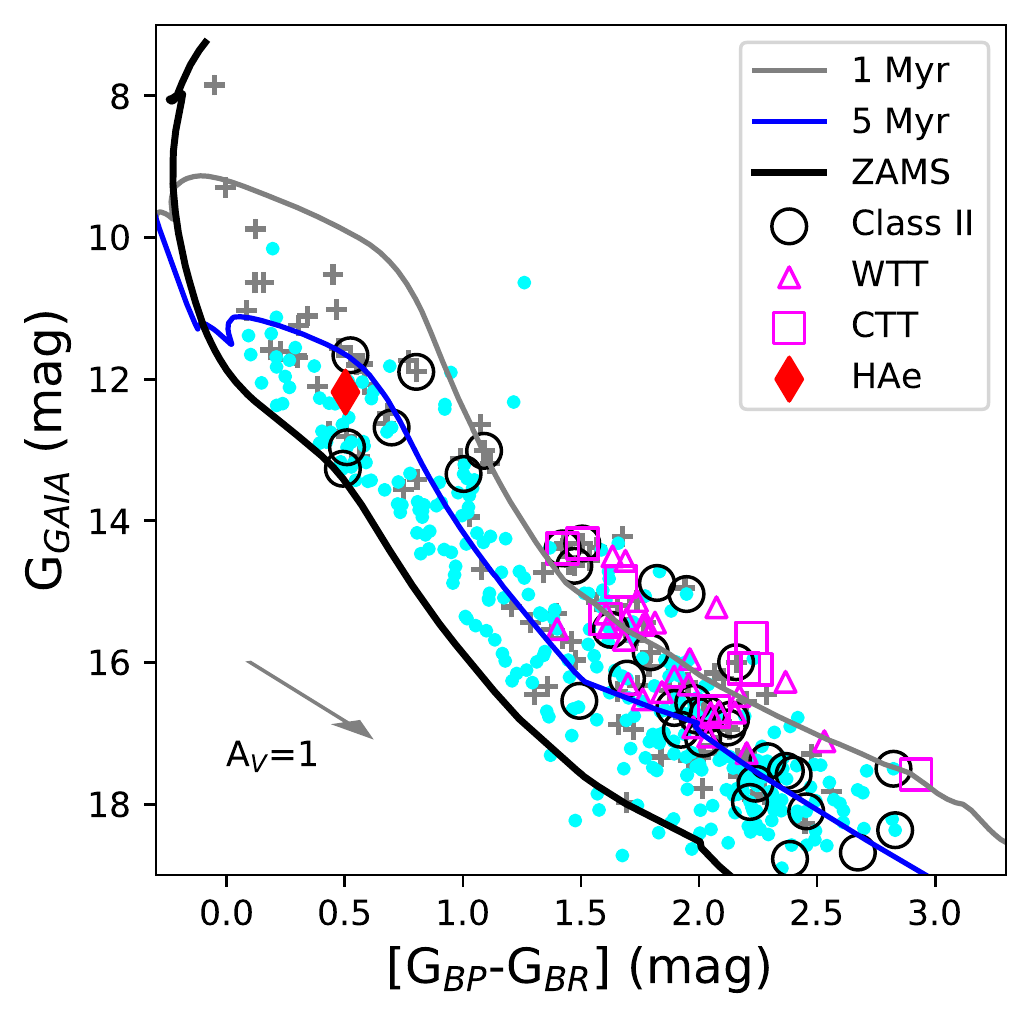}
\caption{Color-magnitude diagram obtained from {\it Gaia}-DR2 photometry compared with isochrones from PARSEC.
The symbols are the same as Fig. \ref{fig:wise_cor},  with open circles to highlight the sources exhibiting IR-excess (Class II).
}
\label{fig:ages}
\end{figure}

The second step for classifying our sample was to determine the range of the stellar ages using {\it Gaia-DR2} photometry.
Figure \ref{fig:ages}  presents the magnitude-color diagram constructed using the G$_{Gaia}$ magnitude as a function of  the [G$_{BP}$] - [G$_{RP}$]  color. In the same plot, we compare the distribution of the points  with isochrones from 
{\it PARSEC} {\footnote{Version v1.2S+COLIBRI PR16 of {\it PARSEC} models \\
available on http://stev.oapd.inaf.it/cgi-bin/cmd.}} \citep{Bressan12, Marigo17}. The models for 1 Myr and 5 Myr were adopted to respectively indicate
the young- and intermediate-ages, expected for the CMa R1 population (see discussion below).
 The 100 Myr  isochrone was chosen to represent the ZAMS ({\textit Zero Age Main 
Sequence}) aiming the comparison with the Main Sequence (MS) colors used by other works discussed in Sect. \ref{sect:3.2}.

Most of the T Tauri stars; other previously known objects, and sources
exhibiting  IR excess (Class II) are $\sim$ 5 Myr or younger. 
About half of the new members and candidates presented in here are  in the same range of ages,
while the other half are older, but still in the PMS phase. 

It is interesting to discuss the spread of ages in comparison with the kinematic characteristics presented  in Sect. \ref{sec:pm}. As previously noted by \citet{GH09},  
CMa R1 has a mixed population. The sources found outside the nebula (to the W direction) tend to be in the  5-10 Myr range (or older), which can be explained by 
their location in a region depleted of gas, where the star formation seems to be ceased. This older group
coincides with the sub-cluster CMa05 located to the NW direction (see Table \ref{tab:pm}) for which SS20 estimated an age of 25$^{+3}_{-7}$ Myr. 

On the other hand, the main clustering associated with the cloud tend to be younger.
For the group CM06, which coincides with our main clustering of  members, SS20 estimate a mean age of 6$^{+1}_{-1}$ Myr. 
The method adopted by those authors is based on the {\it fitCMD} code from \citet{Bonatto19} that provides 
accurate age estimates for stellar clusters. 

In summary, the age estimation obtained here is validated by other method, giving weight to the characterization of our sample as likely PMS members of CMa R1.


\section{Analysis}
\label{sec:sec4}

In this section we describe the methodology we adopted to explore the observational data of our targets, by highlighting some of the results from the literature that have inspired this work.

By comparing the photometric data at different bands, we are able to estimate colors of the sources
in several spectral domains, which characterize the cloud members.
In particular, the flux of H$\alpha$ and Ca II IRT features combined with {\it u', g', r'} colors, give us inferences on
accretion rate and magnetic activity. 

To identify sources showing
color excess, it is necessary to refine the ZP magnitudes (mentioned in Sect. \ref{sec:t80}) by comparing theoretical
data with the observed (\textit {u', g', r', i'} and H$\alpha$) color-color diagrams, as presented in the following subsection

\begin{deluxetable}{lccccccc}
\label{tab:log}
\tablecaption{Details of the {\it T80S} observations.}
\tablewidth{700pt}
\tabletypesize{\scriptsize}
\tablehead{
\colhead{filter} & \colhead{$t_{\rm_{exp}}$} & \colhead{seeing} & \colhead{m$_{sat}$} &\colhead{m$_{50\%}$} &
\colhead{$\lambda$} & \colhead{ZP$_o$} & \colhead{ZP$_f$} \\
\colhead{} & \colhead{s} & \colhead{\arcsec} & \colhead{mag} & \colhead{mag} & \colhead{\AA} & \colhead{mag} & \colhead{mag}
}
\startdata
{\it u'} & 681 & 2.042 & 9.40 & 22.10 & 3574 & 19.64 $\pm$ 0.09 & 18.80 \\
J0378 & 660 & 1.747 & 9.60 &21.83 & 3771 & 19.07 $\pm$ 0.09 & 18.55 \\
J0395 & 354 & 1.763 & 9.87& 21.47 & 3941 & 19.09 $\pm$ 0.11 & 19.00 \\
J0410 & 177 & 1.844 & 10.02& 21.53 & 4094 & 20.10 $\pm$ 0.11 & 20.20 \\
J0430 & 171 & 1.632 & 10.08 & 21.54 & 4292 & 20.26 $\pm$ 0.10 & 20.45 \\
{\it g'} & 99 & 1.772 &10.06 & 21.88 & 4756 & 22.54 $\pm$ 0.09 & 22.74 \\
J0515 & 183 & 1.639 & 9.95 & 21.33 & 5133 & 20.41 $\pm$ 0.08 & 20.65 \\
{\it r'} & 120 & 1.617 & 10.00 & 21.12 & 6260 & 22.46 $\pm$ 0.06 & 22.75 \\
J0660 & 870 & 1.775 & 9.96 & 21.02 & 6614 & 19.95 $\pm$ 0.08 & 20.15 \\
{\it i'} & 138 & 1.620 & 10.01 & 20.54& 7692 & 22.30 $\pm$ 0.06 & 22.44 \\
J0861 & 240 & 1.397 & 10.02 & 20.23 & 8611 & 20.58 $\pm$ 0.07 & 20.70 \\
{\it z'} & 168 & 1.548 & 10.50 & 20.27 & 8783 & 21.75 $\pm$ 0.06 & 21.85 \\
\enddata
\tablecomments{
Seeing  indicates the quality of photometric observations. The saturation limit is given by  m$_{sat}$ and  photometric depth is m$_{50\%}$.
Last columns give the first estimate of Zero Point (ZP$_o$) and the adopted value (ZP$_f$), for which we assume uncertainty of 0.1 mag.
}
\end{deluxetable}

\subsection{Calibration}
\label{sec:calib}

The data calibration and evaluation of ZP magnitudes were based on spectral energy distribution (SED) fitting
and color-color diagram from $g'$-, $r'$-, and $i$-band. We analyzed only sources exhibiting well-determined fluxes
(signal-to-noise ratio S/N $>$ 10) measured with the filter J0660. We also used the
photometric depth of the S-PLUS images presented by \citet[][]{Claudia19}, by selecting only sources 
that are brighter than 21.12 mag for {\it r'}-band, for instance,  leading to a list of about 12,000 sources  that have good photometric quality. 
Table \ref{tab:log} gives the m$_{50\%}$ photometric depth corresponding to the magnitudes at which 50 percent of the total detected sources are included. 
On the other hand, the upper limit of saturation is 10 mag at {\it r'}-band, which corresponds to the brighter source detected in the {\it T80S} image without indication of saturated pixels. In the observed field, there are only four B-type stars brighter than this limit that could be affected by saturation: HD~53396, HD~53456, HD~53457, 
for which no photometry was extracted from the {\it T80S} catalog, and HD~53035 that was detected with {\it r'} = 9.03 mag. Excepting these stars, 
the sources in our sample are fainter than the saturation limit (m$_{sat}$) given in Table \ref{tab:log}.

\begin{figure}
\includegraphics[angle=0,width=\columnwidth]{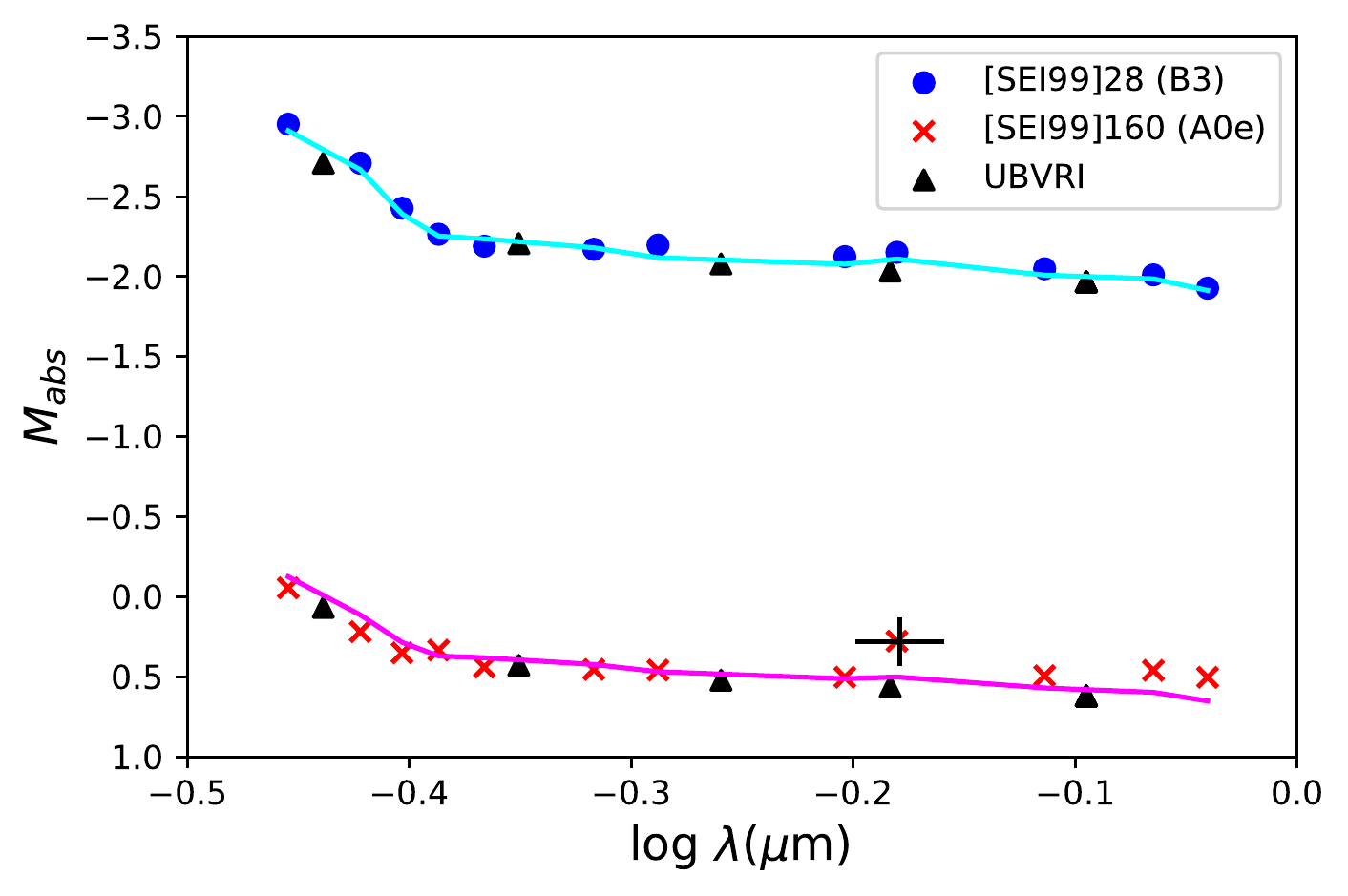}
\caption{Absolute magnitudes ({\it S-PLUS} photometric system) provided by {\it PARSEC} models (ZAMS) for masses: 5 M$_{\odot}$ (cyan line), and 3 M$_{\odot}$ (pink line), compared with the 
 {\it T80S} photometry obtained for two intermediate-mass stars from \citet[][SEI99]{Schev99}. The SED fitting is also confirmed by adopting the  theoretical colors from the {\it UBVRI} system (black triangles)
As it is typical for Ae stars, [SEI99]160 shows excess at H$\alpha$ band, which is indicated by black error-bars.}
\label{fig:sed}
\end{figure}

\begin{figure*}
\includegraphics[angle=0,width=\columnwidth]{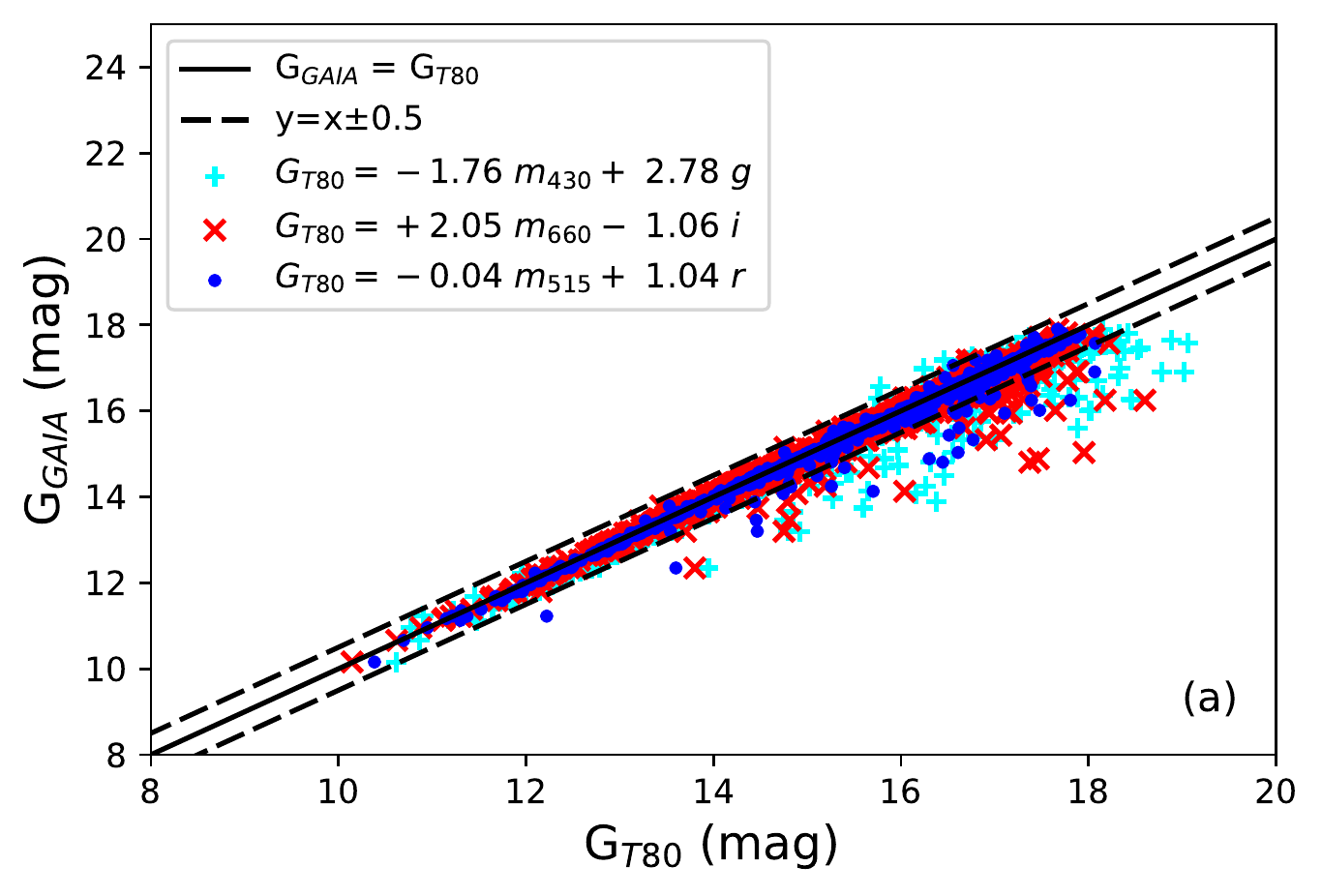}
\includegraphics[angle=0,width=\columnwidth]{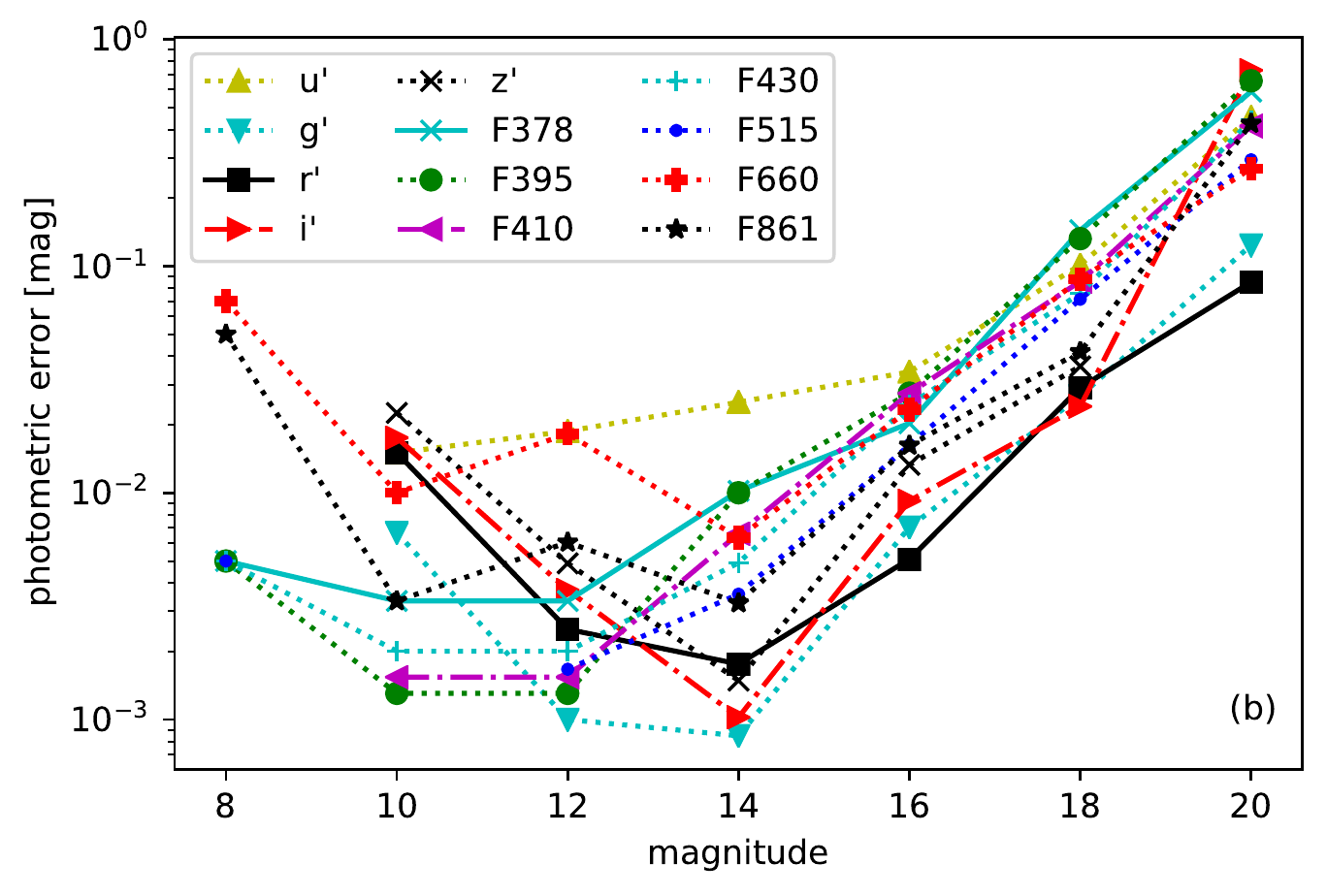}
\caption{(a) Different relations used to transform {\it T80S} photometry into G-band magnitude of the {\it Gaia-DR2} system. 
(b) Distribution of mean photometric errors  as a function of magnitude for each filter,  presented  in 2 mag wide bins: 7-9, 9-11, etc.
}
\label{fig:Gmag}
\end{figure*}

Among the intermediate-mass stars from \citet[][SEI99]{Schev99} that were detected by {\it T80S}, we selected two
to fit their SED: [S99]28 (spectral type B3 ) and [SEI99]160 (HT CMa; A0e), which were used as
calibration stars.
 The SED fitting was obtained by comparison of the observed absolute magnitudes with the theoretical
 values extracted from {\it PARSEC}  model  for the 100 Myr isochrone that was chosen to represent the ZAMS.
Vega magnitudes from two photometric systems were adopted: S-PLUS and {\it UBVRI} \citep{Maiz,Bessell}. 

We found for [SEI99]28 a good match with the 5 M$_{\odot}$ stellar model, while
 the magnitudes of [SEI99]160 are better represented by the 3 M$_{\odot}$ model, excepting for an excess in
 H$\alpha$ noted for this star.

Figure \ref{fig:sed} shows the theoretical SEDs compared with data of calibration stars.
A best fitting was obtained by adopting A$_{V}$ = 0.3 mag ([SEI99]28) and 1.0 mag ([SEI99]160)
for reddening correction of the observed magnitudes. These A$_{V}$ values agree well with the visual extinction map
available for the CMa region \citep{GH08}.
Based on the results of the SED fitting, we obtained the final values for ZP$_f$, which are reported in Table \ref{tab:log}.
 
 An independent validation of the ZP$_f$ values was obtained by following \citet[][see also \citealt{Molino14}]{Claudia19} that give the transformations between magnitudes from {\it Gaia-DR2} and {\it T80S}, expressed by:
  \begin{eqnarray}
G_{T80S} & = & +2.0528 ~ m_{660} - 1.0594 ~i' \label{eq_g_i} \\
G_{T80S} & = & -0.0372 ~  m_{515} + 1.0442 ~  r' \label{eq_g_r} \\
G_{T80S} & = & -1.7642 ~  m_{430} + 2.7824 ~  g' \label{eq_g_g}
\end{eqnarray}
where {\it i', r', g'}, m$_{660}$, m$_{515}$, and m$_{430}$ are the apparent magnitudes measured in the respective bands and calibrated with the ZP$_f$ given in Table \ref{tab:log}.
 
Figure \ref{fig:Gmag}a shows the empirical G-band magnitudes (G$_{T80S}$) obtained by a linear fit of magnitudes from a set of {\it T80S} filters 
and the {\it Gaia-DR2} G-band magnitudes (G$_{Gaia}$). 
A good correlation of G$_{T80}$ with G$_{Gaia}$ is obtained for 
most of the sources, which are found in between the dashed lines  representing 0.5 mag deviation.  
The faint sources (G$_{T80S} >$ 16 mag) show a dispersion, possibly due to different reasons
(e.g. variability;  photometric errors; etc).  Figure \ref{fig:Gmag}b displays the distribution of mean values of errors as a function of observed magnitudes
for each {\it T80S} filter, showing  that sources fainter than 16 mag have errors larger than 0.05 mag. Bright sources ($\sim$ 8 mag) tend to show   larger errors at filters F660 and F861 than those measured in other bands, probably due to nebular contamination combined with saturation effects.

Despite the dispersion of faint sources, the result shown in Fig. \ref{fig:Gmag} can be considered as a validation of the ZP calibration adopted by us. Additional criteria confirming this hypothesis are discussed as follows.


\subsection{Accretion}
\label{sect:3.2}

\citet{Venuti14} used {\it u, g,} and {\it r} photometry to estimate
accretion rates for the pre-main sequence population of the young cluster NGC~2264. The candidates showing
UV-excess were selected from the
$[u-g] \times [g-r]$ and $[r] \times [u-r]$ diagrams,  where the Classical T Tauri stars (CTTs) occupy a
distinguished locus, separated from weak-line T Tauri (WTTs) and  field stars.

The $[u-g]$ color is also useful to distinguish candidates showing evidences of UV-excess due to accretion process.
For instance, \citet{Kalari15}  obtained H$\alpha$ and {\it u}-band accretion rates for CTTs in the Lagoon Nebula M8 by
using $ugri$ and H$\alpha$ photometry from the VST Photometric H$\alpha$ survey \citep[VPHAS+,][]{Drew14}.

\begin{figure*}
\center{
\includegraphics[angle=0,width=\columnwidth]{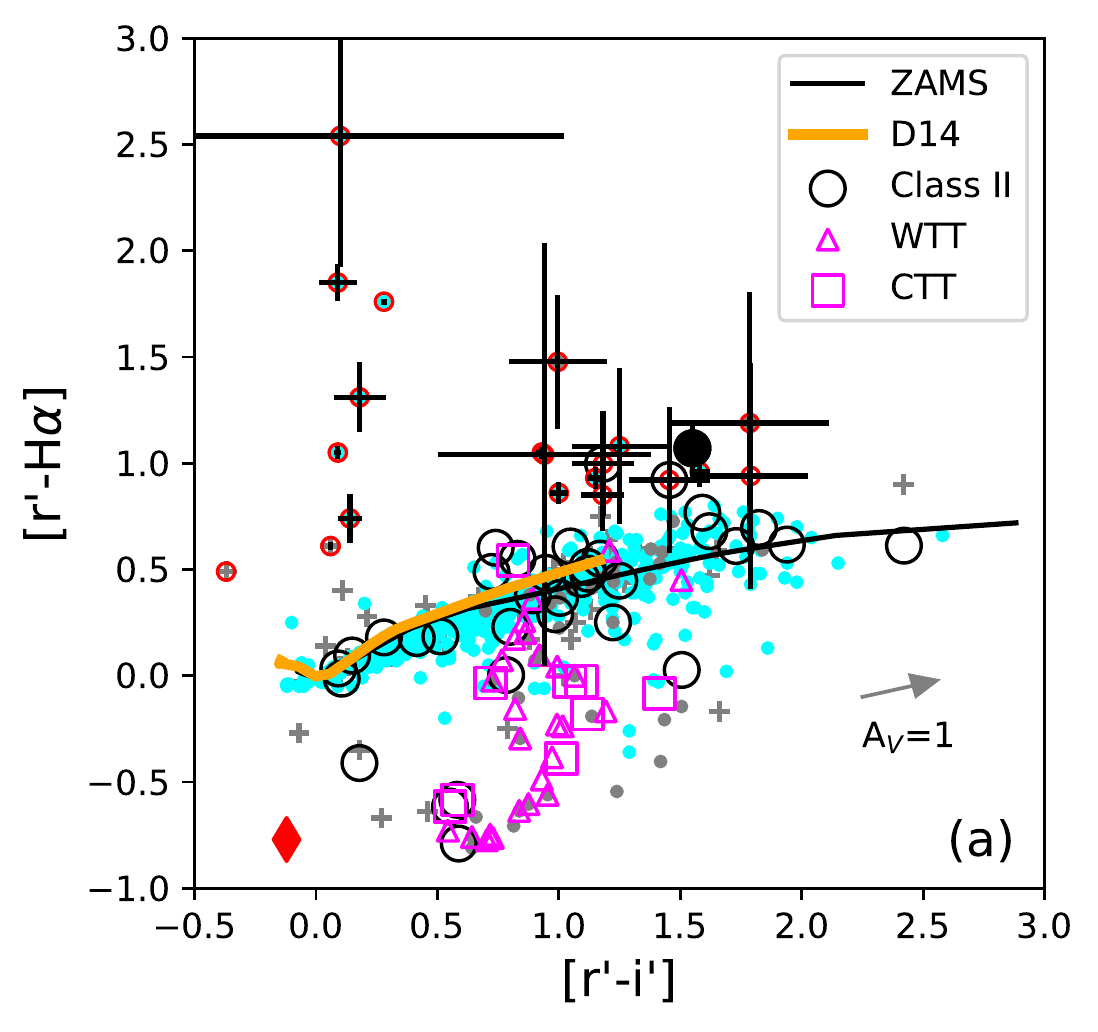}
\includegraphics[angle=0,width=8cm]{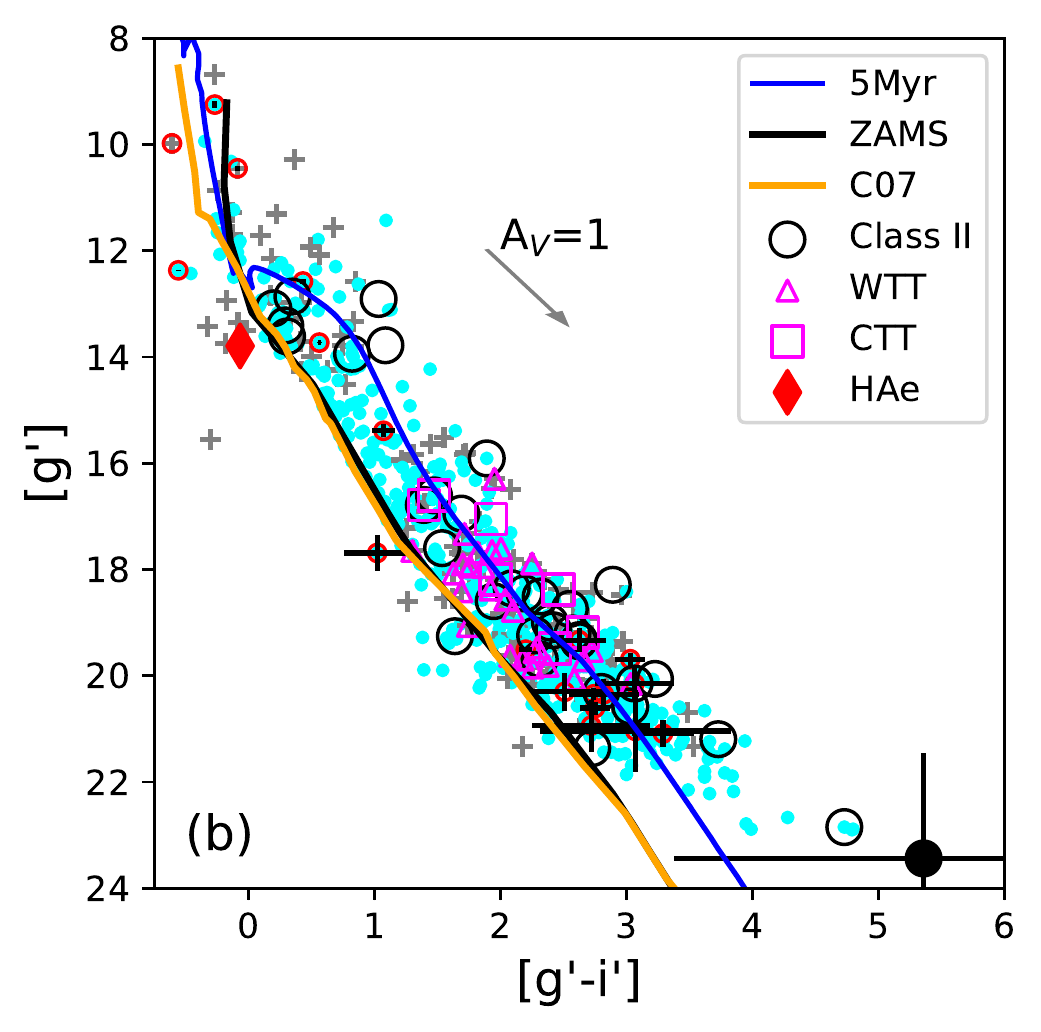}
\includegraphics[angle=0,width=\columnwidth]{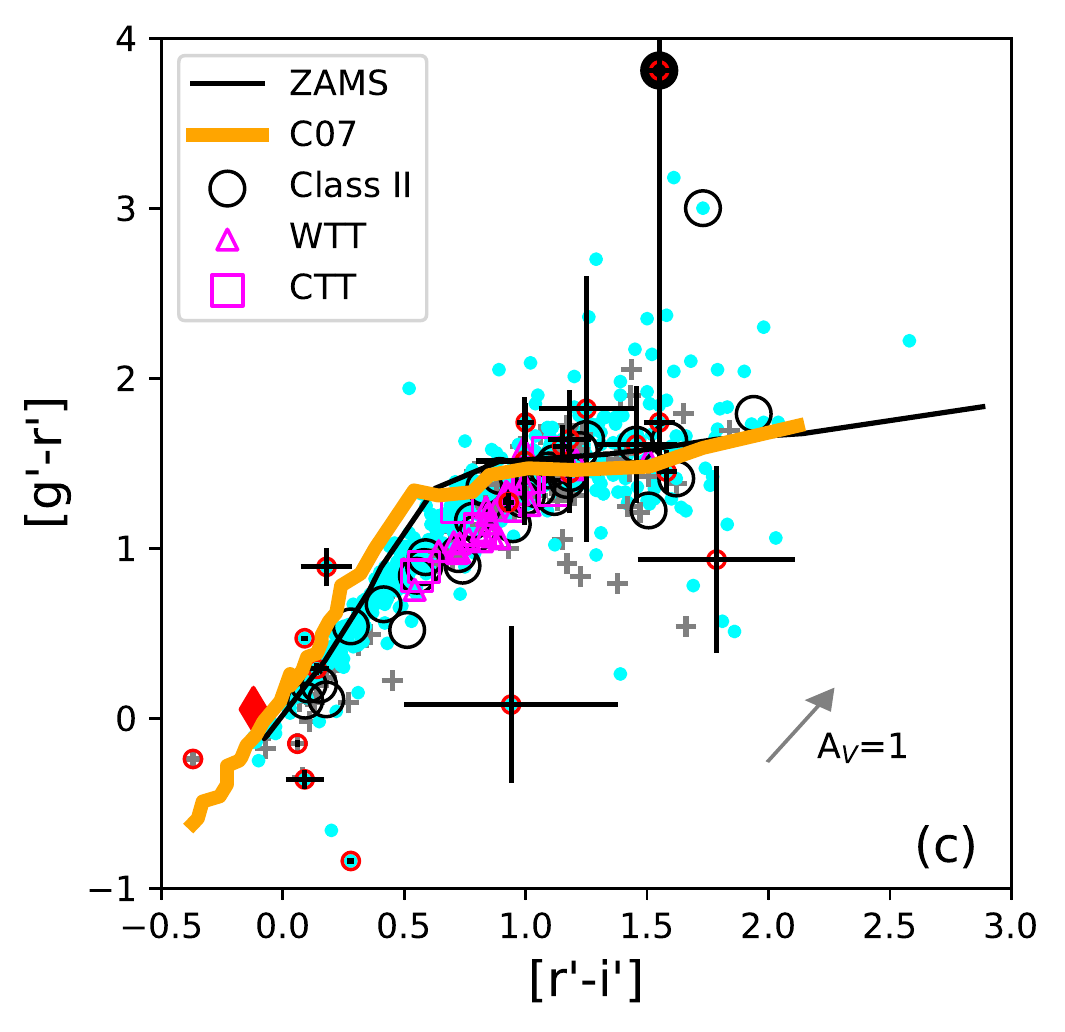}
\includegraphics[angle=0,width=\columnwidth]{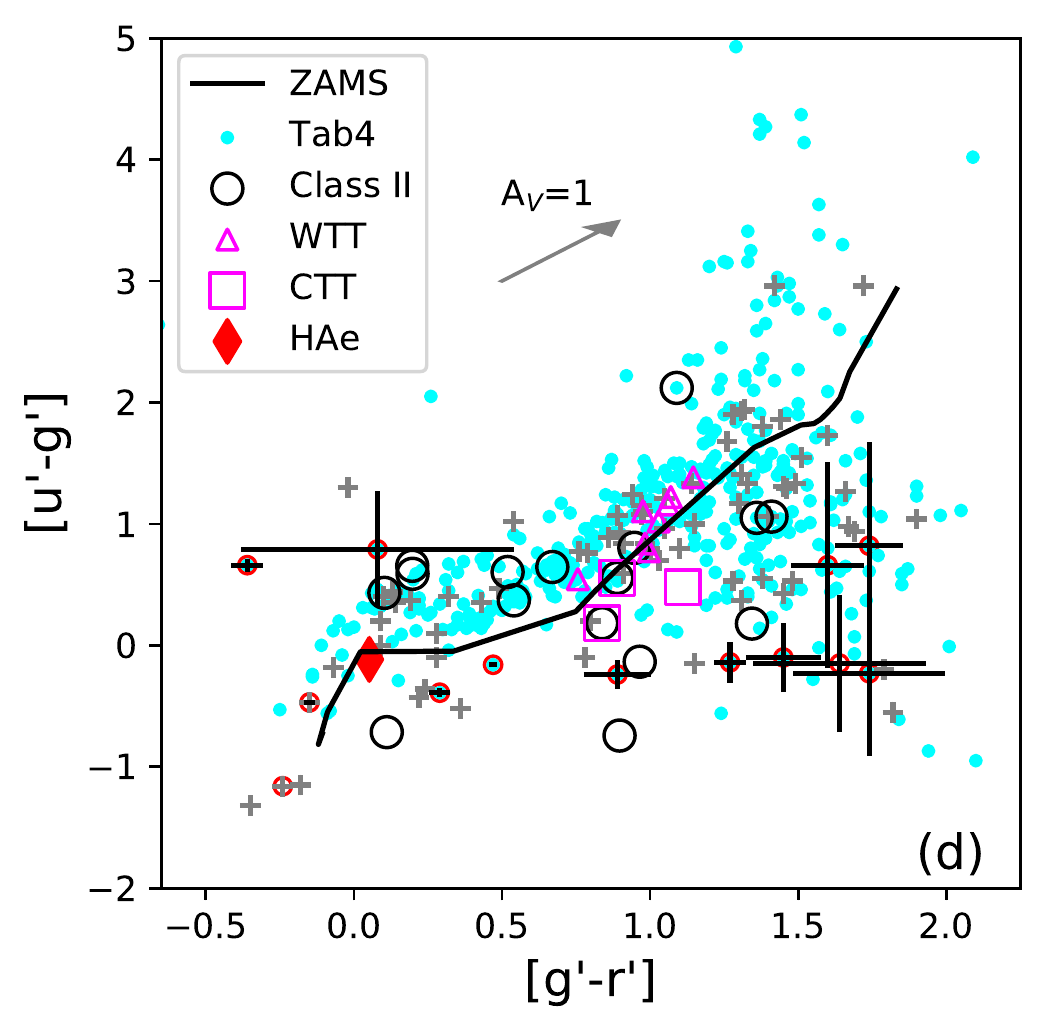}
\caption{Color-color diagrams obtained for {\it T80S} data compared with theoretical colors from
{\it PARSEC} model (ZAMS, black line), and the Main Sequence (orange line) from \citet[][D14]{Drew14}, or
 \citet[][C07]{Covey07}.  The previously known candidates (grey $+$), and the confirmed PMS members \citep{Fernandes15}
 T Tauri stars (pink squares and triangles)  and Herbig Ae star (red diamond) are  listed in Table \ref{tab:lista6}. 
 The new members  and candidates revealed in this work (cyan dots) are listed in Tables \ref{tab:lista7} and \ref{tab:lista8}.
 Open black circles indicate the objects showing {\it WISE} colors typical of Class II stars (see Fig. \ref{fig:wise_cor}), and a filled circle shows one Class I candidate. 
 In panel (a) we note  22 objects  exhibiting  $[r'-H\alpha]$ color excess, which are marked by red circles with error-bars (for some of them the errors are lower than the symbol size).
 }
\label{fig:ugri}
}
\end{figure*}

Following \citet{Venuti14} and \citet{Kalari15}, we present our results in the 
diagrams shown in Fig. \ref{fig:ugri}, by comparing the observed colors
with theoretical  MS and  ZAMS  \citep[{\it PARSEC} model,][]{Bressan12}, 
whose fitting may validate the ZP calibration.
 The previously known candidates  exhibiting good quality photometry (see Sect. \ref{sec:calib}) are indicated by grey symbols, while
cyan dots are used to show the new members and candidates.

We also highlight  in
Fig. \ref{fig:ugri} the distribution of PMS stars characterized  on the basis  of {\it GMOS} spectra \citep{Fernandes15}.

In Fig. \ref{fig:ugri}a, we compare $[r'-H\alpha] \times [r'-i']$  with MS colors given by  \citet{Drew14}.
According to \citet{Kalari15} accreting stars are expected to be found above the MS line,  in a  locus where bluer sources ($[r'-i'] <$ 1 mag)  
have $[r'-H\alpha] \gtrsim$ 0.5 mag, and redder sources have $[r'-H\alpha] \gtrsim$ 0.8 mag. The presence of objects at that region of the plot is 
interpreted as an evidence of H$\alpha$ excess related to accretion activity.

 Here, we draw attention to  22 objects  (marked with error bars) that appear with H$\alpha$ excess in Fig. \ref{fig:ugri}a: six are known 
candidates  and 16 are new  members. Among them there  are two Class II stars (black circles, see Sect. \ref{sec:class}), and one Class I
 candidate (highlighted by a filled circle). However, no $H\alpha$ excess is clearly seen for the other sources having IR excess. 
Furthermore, none of the confirmed PMS stars can be used as indicator of accretion, since they are found below the MS. 
This is expected for the WTT  stars, but it should not be the case for the HAe  and the CTT  stars. We tentatively suggest that 
it can be due to the low S/N photometry acquired for these faint sources (see Sect. \ref{sec:pm}).
 
Among the other candidates previously known  that appear 
in the bottom of Fig. \ref{fig:ugri}a there  are two B-type stars 
listed by \citet{Schev99}: [SEI99]102 and [SEI99]111, for the latter no H$\alpha$ emission was detected in previous studies 
\citep{Cohen79,Wira86, Schev99} or in the present work.  The other stars located at  this region of the plot are X-ray sources that could be WTTs, which typically do not exhibit strong H$\alpha$ emission.

 Figure \ref{fig:ugri}b is a color-magnitude diagram  comparing the distribution of our sample and the MS line  from \citet[][]{Covey07}, by adopting the $[g'] \times [g'-i']$ magnitudes. The data points are also compared with the ZAMS that coincides with the MS in this plot, 
and the 5 Myr isochrone ({\it PARSEC}), representing a mean value of  age estimated for the sample (see Fig. \ref{fig:ages}).

The observed  $[g'-r'] \times [r'-i']$ colors compared with  MS from  \cite{Covey07} are presented in   Fig. \ref{fig:ugri}c.
In a similar study, results from  \citet{Venuti14} show no-accreting stars (WTTs) coinciding with the MS, while CTTs are 
 widely scattered around the MS. 
 In our case, only a few objects are found below the MS curve, but we cannot be sure if this is related to accretion or not, since these stars 
 may not exhibit  {\it u'}-band excess. 

 Figure \ref{fig:ugri}d presents the $[u'-g'] \times [g'-r']$ diagram used to verify the occurrence 
of UV-excess in our sample. We also compare our data with the ZAMS.
Despite the large dispersion, probably due to low quality of photometry in the {\it u'}-band, most of the objects
are likely following the
theoretical ZAMS. Only a few stars are found below this curve, which is the expected locus of stars showing UV-excess \citep{Kalari15}.

By considering only objects of our sample,  we suggest that 3.2$\pm$0.7 percent (22/694)
seem to have accretion activity as indicated by both  $[r'-H\alpha] \times [r'-i']$  and $[u'-g'] \times [g'-r']$ diagrams,
which confirm our previous results about the lack of disk-bearing stars associated with Sh~2-296, which is the main nebula of CMa OB1/R1.

\begin{figure}
\includegraphics[angle=0,width=\columnwidth]{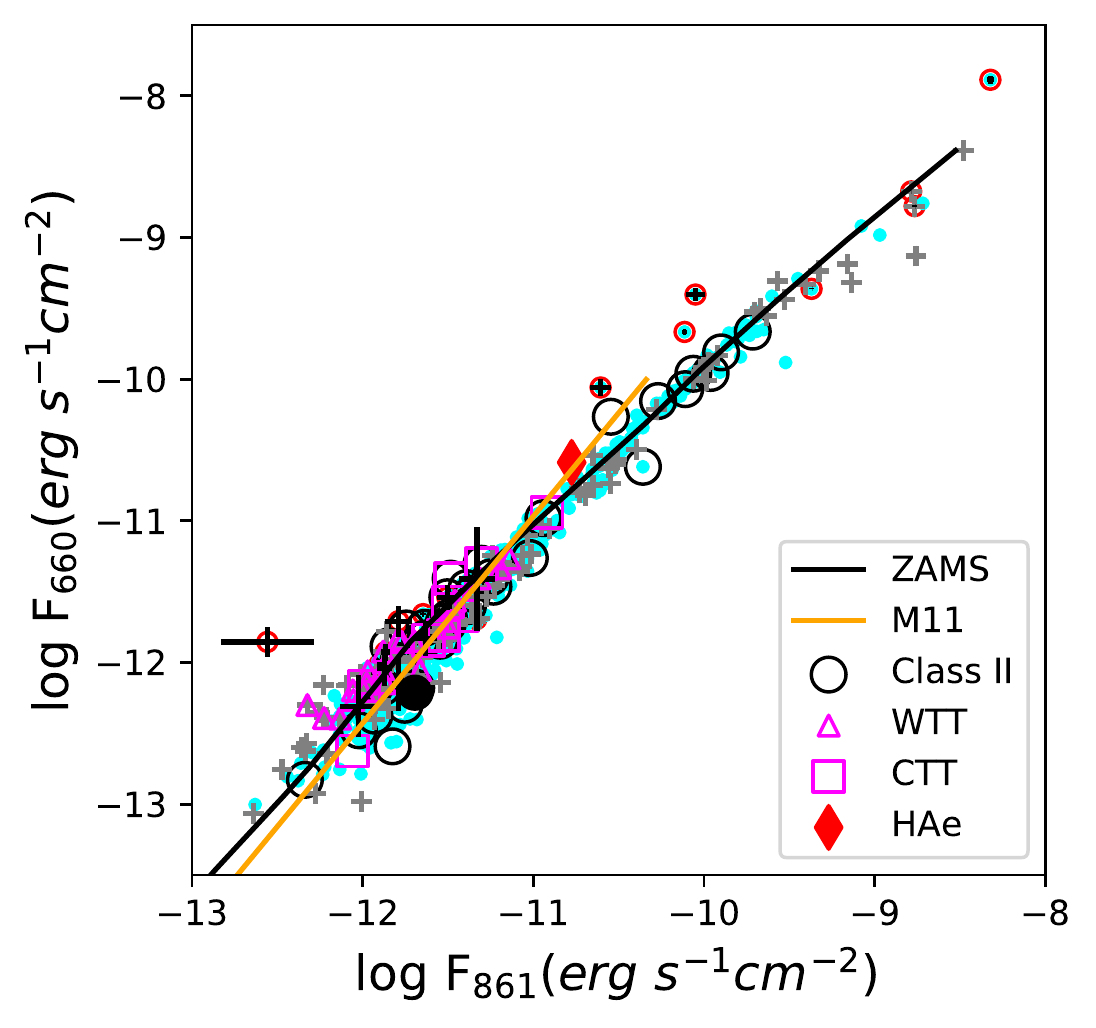}
\caption{Broad band fluxes associated with the
H$\alpha$ and Ca triplet emission, respectively estimated from the m$_{660}$ and m$_{861}$ magnitudes observed with {\it T80S}. 
The flux-flux relation is comparable with results from \citet[][M11]{Martinez11}, where active late-type stars are expected to appear above the MS line. Symbols are the same as Fig. \ref{fig:ugri}.  }
\label{fig:haca}
\end{figure}


\subsection{Magnectic Activity}

Due to the lack of spectroscopic observations for our sample, we can not evaluate  the spectral flux to probe magnetospheric accretion, nor to estimate accretion rates.
Alternatively, we can obtain an indirect evidence of activity by using photometry taken at spectral bands containing the aimed spectral lines.

The measurements of H$\alpha$ flux, combined with the flux of   Ca II  lines is commonly used as indicators of chromospheric activity
in late-type dwarf stars \citep[see references given by][]{Martinez11}.
The flux-flux relationship between H$\alpha$ and Ca II IRT ($\lambda$ 8498 \AA) evaluated by those authors shows that young stars
have an excess of flux at these bands, when compared with the empirical linear fit found for F- G- and K-type active stars. Despite the differences on the
considered wavelength, we used magnitudes measured with filters J0660 and J0861 (see Table \ref{tab:log}), aiming to roughly reproduce the flux-flux
relationship between  H$\alpha$ and Ca II IRT, respectively. 
 It is important to note that we are not estimating line fluxes here, but just comparing broad band 
fluxes\footnote {The conversion of instrumental magnitudes (m$_{\lambda}$) to flux (F$_{\lambda}$) is given by
$F_{\lambda} (Jy) = 3631 \times 10^{-0.4[m_{\lambda} + ({\rm ZP}_f)_{\lambda}] }$}, with no subtraction of continuum.

Figure \ref{fig:haca} shows the flux-flux relationship  
between the fluxes $F_{660}$ and $F_{861}$ observed with {\it T80S}, showing  a few sources above the MS  line that
could indicate H$\alpha$ excess.
This is not unexpected since  most of the objects studied by \citet{Fernandes15} are WTTs, and the
A- and B-emission stars \citep{Schev99} present in our sample are not strong  H$\alpha$ emitters.


\section{Conclusions}
\label{sec:sec5}
\subsection{Summary of results}

We used the \textit{Gaia}-DR2 astrometric data to evaluate the membership probability (P) and identify the CMa R1 members based on  proper motion.
From  \textit{Gaia}-DR2  we extracted only the objects showing parallax $\varpi$ = 0.8 - 1.25 mas, compatible with the
distance of the cloud \citep[d $\sim$ 1 kpc,][]{GH08}, RUWE $<$ 1.4 and  $\varpi / \sigma_{\varpi} > 3$
resulting in a list of  $\sim$ 4,200 objects in the area observed with {\it T80S}. 

Our {\it T80S} observations covered a $\sim$ 2 deg$^2$  area around Sh~2-296, the main nebula of the CMa R1 region, where more than 70 thousand objects were 
detected,  17 percent of them exhibiting good quality photometry.

 We selected 669 sources on the basis of membership probability that was estimated with the Bayesian technique. The mean values of proper motion for 523 probable members  (P$\geq$50 percent) are: 
( $\mu_{\alpha} \cos \delta \sim$ -4.1 $\pm$0.6, $\mu_{\delta} \sim$ 1.5 $\pm$0.4) mas yr$^{-1}$. 
We also considered 146  candidates (P $<$ 50 percent)
exhibiting proper motion compatible within a  3$\sigma$-interval from the mean values.
 The other objects extracted from {\it Gaia} are very likely field-stars (membership probability: P $<$ 1 percent).
 
Table \ref{tab:lista2} contains 155 stars that are candidates  previously identified in the literature, but only 149   were included  in our final sample because six sources were not detected by {\it T80S}. Other 31 previously known objects included in our sample are spectroscopically confirmed PMS stars (Table \ref{tab:lista3}),
but they have low S/N photometry. 
In total, our final sample contains 694 sources, 180 of them are previously known objects.
Therefore, the membership analysis based on \textit{Gaia}-DR2 data revealed  395 new members (not previously identified in the literature) and 119 new candidates. 
Table \ref{tab:lista4} and Table \ref{tab:lista5} give   the kinematic data and identification of these new members and candidates, respectively.

The color-magnitude diagram  G$_{Gaia}$  $\times ~[G_{BP}-G_{RP}]$ was used to estimate the age of the sources. A mean value of 5 Myr was found, which
characterizes our sample as likely PMS members of CMaR1. The presence of IR-excess was investigated on the basis of {\it WISE} colors that revealed 43 objects: 2
are Class I  and 42 are Class II.  Only six of the T Tauri stars have IR-colors typical of Class II (none is Class I), which indicates that most of the confirmed PMS stars
of our sample are Class III, exhibiting low IR emission.

The photometric zero-points (ZPs) and calibration of the {\it T80S} data were obtained by using different methods (see Sect. \ref{sec:calib}). First, we performed the SED fitting of two stars with well determined spectral type, by adopting \textit{PARSEC} evolutionary models.

The ZPs were also evaluated by analyzing the $g', r', i'$ colors diagram and correlations between the G-band magnitude from \textit{Gaia}-DR2 ($G_{Gaia}$) and empirical values
obtained from {\it T80S} data ($G_{T80}$). 

Table \ref{tab:lista6} gives the  photometric data for previously known objects that were observed by {\it T80S}.
The photometric data for the new members and candidates 
are given in Tables \ref{tab:lista7} and \ref{tab:lista8}, respectively. 

Color-color diagrams were used to confirm the occurrence of CMa R1 members showing color excess related to accretion process,
which did not reveal more than 22 stars with excess in the H$\alpha$ or the \textit{u'}-band. 

Magnetic activity was investigated according the flux-flux relations used by \citet{Martinez11} comparing H$\alpha$ and  Ca II H \& K lines fluxes typically found for late-type stars showing chromospheric activity. An excellent correlation between the broad band F$_{660}$ and F$_{861}$ fluxes is found, but only a few stars 
have indications of H$\alpha$ excess likely due to accretion.

\subsection{Concluding remarks}

Previously we identified a few hundred young stars candidates associated with the nebulae in CMa R1, but our studies were restricted to small areas, such as the fields covered by \textit{XMM-Newton} observations. 
Aiming to explore a more complete  and reliable sample of candidates,  the {\it Gaia}-DR2 catalog was used to select the kinematic members, for which
we analyzed multi-band {\it T80S} imaging observations that provided 
very good indicators of accretion activity. 
 
Despite a large sample of new PMS candidates were identified,  a small fraction of them show signs of accretion processes and/or magnetic activity. 
Only  3.2$\pm$0.7 percent of our sample seem to have H$\alpha$ and/or UV excess, 
which gives an estimation of accreting fraction.

Based on  the number of Classes I and II sources (43/694) we estimate a disk fraction of 6.2$\pm$0.9 percent. We have noticed
in our sample the presence of sources that may be part of a sub-group, which was identified in the NW direction of the region. 
But none of these sources have IR-excess. Therefore, the number of disk-bearing stars is evaluated here for the whole population.

Comparing CMa R1 with other regions, we note that our results are lower than the values expected for young stellar groups with similar ages  ($\sim$ 5 Myr). 
For instance,  a disk fraction of more than 10 percent is reported for NGC~2362 \citep{Haisch01};  and 20 percent for Upper Sco \citep{Fedele10, Cloutier14}. 
The accretion fraction estimated for these regions is $\sim$ 5 percent \citep{Fedele10}. 
More recently, \citet{Briceno19} presented the decline of the accreting fraction, as well as the percentage of disk-bearing stars, as a function of the mean 
age of each sub-group of the  Orion OB1
star-forming region. For Orion$_{1b}$ sub-group that is 5 Myr old, they report 
 an accreting fraction of about 10 percent, and a disk fraction of $\sim$ 15 percent.

We conclude that analyzing the optical photometric data  gave us a confirmation of  our previous results, which were based on spectroscopy and  near-infrared data.
Despite the use of a much large sample of candidates, the  absence of disk-bearing stars in the region is confirmed. Somehow, circumstellar accretion has finished  for the young stars in this region, probably due to 
the star formation scenario where at least three supernova events occurred a few millions years ago. 
This could be the cause of the disks disappearance earlier than usually is found  in other star-forming regions at same age.

\begin{longrotatetable}
\begin{deluxetable*}{cccccccccc}
\label{tab:lista2}
\tablecaption{Objects identified in the literature, giving membership probability
(P\%), parallax ($\varpi$),  proper motion
($\mu_{\alpha \cos \delta}$ and $\mu_{\delta}$), and other data from {\it Gaia}-DR2.}
\tablewidth{700pt}
\tabletypesize{\scriptsize}
\tablehead{
\colhead{\textit{Gaia-DR2}} & \colhead{2MASS} & \colhead{RA} & \colhead{Dec} & 
\colhead{P} & \colhead{$\varpi$} & 
\colhead{$\mu_{\alpha} \cos \delta$} & \colhead{$\mu_{\delta}$} & \colhead{G} & 
\colhead{Note} \\
\colhead{} & \colhead{} & \colhead{degree} & \colhead{degree} & \colhead{\%} & \colhead{mas} & 
\colhead{mas yr$^{-1}$} & \colhead{mas yr$^{-1}$} & \colhead{mag} & \colhead{} }
\startdata							
3046091622096649600 & 07013441-1117375 & 105.3934 & -11.2938 & 96 & 0.912$\pm$0.042 & -4.110$\pm$0.072 & 1.459$\pm$0.067 & 15.3 & W011a  \\
3046093301427017600 & 07013774-1112371 & 105.4072 & -11.2104 & 81 & 0.89$\pm$0.21 & -3.72$\pm$0.41 & 2.01$\pm$0.39 & 15.77 &  W096 \\
3046095195509364608 & 07013843-1110147 & 105.4102 & -11.1708 & 0 & 1.104$\pm$0.107 & 2.395$\pm$0.199 & -6.209$\pm$0.190 & 17.33 & W076  \\
3046063893787827840 & 07014236-1123337 & 105.4265 & -11.3927 & 94 & 0.986$\pm$0.088 & -4.354$\pm$0.143 & 1.762$\pm$0.167 & 16.94 & W021  \\
3046088937740213888 & 07014669-1118126 & 105.4445 & -11.3035 & 86 & 0.87$\pm$0.15 & -3.48$\pm$0.27 & 1.39$\pm$0.26 & 17.74 &  W038 \\
3046088976396774016 & 07015002-1118275 & 105.4584 & -11.3077 & 54 & 0.805$\pm$0.028 & -3.569$\pm$0.048 & 2.023$\pm$0.048 & 13.95 & W047a  \\
3046089148195449472 & 07015058-1116405 & 105.4608 & -11.278 & 89 & 0.975$\pm$0.048 & -4.044$\pm$0.087 & 1.928$\pm$0.083 & 15.54 & W005  \\
3046088770238881024 & 07015256-1119089 & 105.469 & -11.3192 & 95 & 0.912$\pm$0.026 & -4.060$\pm$0.046 & 1.694$\pm$0.048 & 13.41 & W016  \\
\enddata
\tablecomments{
Objects from \citet[][]{Fernandes15} are indicated by letter F.  X-ray sources are denoted by letters E, W, C, and S 
corresponding to fields East, West, Center and South (see Fig. \ref{fig:wise}) observed by {\it XMM-Newton} \citep[][]{Santos18}. 
Other identifications and  Spectral Type were compiled by \citet{GH08} mainly  from \citet[][C74]{Claria74} and  \citet[][SEI99]{Schev99}.
The six objects in the bottom lines were not found in the {\it T80S} catalog.
}
\end{deluxetable*}
\end{longrotatetable}


\begin{longrotatetable}
\begin{deluxetable*}{ccccccccc}
\label{tab:lista3}
\tablecaption {The same as Table \ref{tab:lista2}, with the list of confirmed PMS stars \citep{Fernandes15}.}
\tablewidth{700pt}
\tabletypesize{\scriptsize}
\tablehead{
\colhead{\textit{Gaia-DR2}} & \colhead{2MASS} & \colhead{RA} & \colhead{Dec} & 
\colhead{$\varpi$} & 
\colhead{$\mu_{\alpha} \cos \delta$} & \colhead{$\mu_{\delta}$} & \colhead{G} &  \colhead{Notes} \\
\colhead{} & \colhead{} & \colhead{degree} & \colhead{degree} & \colhead{mas} &
\colhead{mas yr$^{-1}$} & \colhead{mas yr$^{-1}$} & \colhead{mag} & \colhead{}}
\startdata
3046019531067048320   &       07035338-1134504  &      105.9724 &     -11.5807  &            0.83 $\pm$	0.04	&    -4.59	$\pm$	0.07	&	1.62	$\pm$ 0.08	&	15.38	&	F10b (CTT)	\\
3045831862476398336	&	07035542-1135149	&	105.9809	&	-11.5875	&		0.84	$\pm$	0.09	&	-3.83	$\pm$	0.15	&	0.47	$\pm$	0.15	&	16.10	&	F18	(CTT?)	\\
3045831862476390144	&	07035666-1134553  &	105.9861	&	-11.5820	&		0.87	$\pm$	0.07	&	-4.34	$\pm$	0.12	&	1.09	$\pm$	0.12	&	16.08	&	F23	(CTT)	\\
3045837772351364864	&	07040041-1133596	&	106.0017	&	-11.5666	&		0.68	$\pm$	0.13	&	-4.74	$\pm$	0.22	&	1.12	$\pm$	0.25	&	15.65	&	F30	(CTT)	\\
3046032179749669888	&	07040314-1123275	&	106.0131	&	-11.3910	&		0.50	$\pm$	0.27	&	-5.43	$\pm$	0.51	&	3.86	$\pm$	0.47	&	17.58	&	F36c	 (CTT)	\\
3046025582680267776	&	07040290-1132074	&	106.0121	&	-11.5354	&		1.38	$\pm$	0.27	&	-5.11	$\pm$	0.49	&	1.51	$\pm$	0.47	&	16.69	&	F37	(CTT)	\\
3045838219027859456	&	07042625-1131207	&	106.1094	&	-11.5224	&			\nodata		&		\nodata		&		\nodata		&	14.86	&	F60	(CTT?)	\\
3046027536887737984	&	07040225-1125429	&	106.0094	&	-11.4286	&		0.86	$\pm$	0.04	&	-4.46	$\pm$	0.09	&	1.43	$\pm$	0.10	&	12.18	&	F34a	 (HAe)	\\
\enddata
\tablecomments{
Identification and classification of Classical T Tauri (CTT), weak-line T Tauri (WTT), and Herbig Ae (HAe) stars were obtained by \cite{Fernandes15} based on {\it GMOS} spectroscopy.
}
\end{deluxetable*}
\end{longrotatetable}

\begin{deluxetable*}{lcccccccc}
\label{tab:lista4}
\tablecaption{ New members of CMa R1  identified in the present work. The astrometric data are the same as Table \ref{tab:lista2} .}
\tablewidth{700pt}
\tabletypesize{\scriptsize}
\tablehead{
\colhead{Gaia-DR2} & \colhead{2MASS} & \colhead{RA} & \colhead{Dec} & \colhead{P} & \colhead{$\varpi$} &
\colhead{$\mu_{\alpha} \cos \delta$} & \colhead{$\mu_{\delta}$} & \colhead{G}  \\
\colhead{} & \colhead{} & \colhead{degree} & \colhead{degree} & \colhead{\%} & \colhead{mas} &
\colhead{mas yr$^{-1}$} & \colhead{mas yr$^{-1}$} & \colhead{mag.} 
}
\startdata
3046062725556787200 & 00000000-0000000 & 105.3028 & -11.4309 & 91 & 0.94$\pm$0.10 & -3.68$\pm$0.15 & 1.59$\pm$0.21 & 15.69  \\
3046107977332044032 & 07011780-1107209 & 105.3242 & -11.1225 & 70 & 1.10$\pm$0.02 & -5.02$\pm$0.04 & 1.94$\pm$0.04 & 13.79  \\ 
3046013792988609408 & 07012500-1132538 & 105.3542 & -11.5483 & 84 & 0.83$\pm$0.16 & -4.69$\pm$0.24 & 2.05$\pm$0.24 & 18.05  \\
3046064958939721728 & 07012916-1121307 & 105.3715 & -11.3585 & 94 & 0.83$\pm$0.05 & -3.78$\pm$0.09 & 1.43$\pm$0.09 & 15.90  \\ 
3046063309672312960 & 07012943-1125450 & 105.3726 & -11.4292 & 93 & 0.92$\pm$0.12 & -3.75$\pm$0.15 & 1.44$\pm$0.18 & 17.02  \\ 
.&. & . & . & . & . & . & . & . \\
.&. & . & . & . & . & . & . & . \\
.&. & . & . & . & . & . & . & . \\
\enddata
\tablecomments{
For illustration, only a part of the table is presented. The entire list of objects is available on-line.
}
\end{deluxetable*}

\begin{deluxetable*}{lcccccccc}
\label{tab:lista5}
\tablecaption{ New candidates (P$<$ 50 percent) identified in the present work. The astrometric data are the same as Table \ref{tab:lista2}.}
\tablewidth{700pt}
\tabletypesize{\scriptsize}
\tablehead{
\colhead{Gaia-DR2} & \colhead{2MASS} & \colhead{RA} & \colhead{Dec} & \colhead{P} & \colhead{$\varpi$} &
\colhead{$\mu_{\alpha} \cos \delta$} & \colhead{$\mu_{\delta}$} & \colhead{G}  \\
\colhead{} & \colhead{} & \colhead{degree} & \colhead{degree} & \colhead{\%} & \colhead{mas} &
\colhead{mas yr$^{-1}$} & \colhead{mas yr$^{-1}$} & \colhead{mag.} 
}
\startdata
3046067364121443840 & 07011449-1123185 & 105.3104 & -11.3885 & 25 & 0.80$\pm$0.04 & -3.04$\pm$0.06 & 1.05$\pm$0.06 & 14.78  \\ 
3046092820387743360 & 07011607-1115513 & 105.3170 & -11.2643 & 34 & 0.96$\pm$0.28 & -3.37$\pm$0.50 & 0.43$\pm$0.44 & 18.78  \\ 
3046067879517494272 & 07011974-1121119 & 105.3323 & -11.3533 & 2 & 0.86$\pm$0.11 & -3.34$\pm$0.19 & 0.39$\pm$0.20 & 16.88 \\
3046092515449873024 & 07012008-1117448 & 105.3337 & -11.2958 & 27 & 0.80$\pm$0.04 & -3.06$\pm$0.06 & 1.05$\pm$0.06 & 14.74  \\ 
3046091759535625856 & 07012338-1118004 & 105.3474 & -11.3001 & 33 & 0.81$\pm$0.04 & -3.15$\pm$0.08 & 1.02$\pm$0.07 & 11.94  \\ 
.&. & . & . & . & . & . & . & . \\
.&. & . & . & . & . & . & . & . \\
.&. & . & . & . & . & . & . & .  \\
\enddata
\tablecomments{
For illustration, only a part of the table is presented. The entire list of objects is available on-line.
}
\end{deluxetable*}

\begin{longrotatetable}
\begin{deluxetable*}{lcccccccccccc}
\label{tab:lista6}
\tablecaption{Photometric data for {\it T80S} sources that were previously identified in the literature (see Tables \ref{tab:lista2} and \ref{tab:lista3}).}
\tablewidth{700pt}
\tabletypesize{\scriptsize}
\tablehead{
\colhead{ID} &\colhead{\textit{u'}} & \colhead{m$_{378}$} &  \colhead{m$_{395}$} & \colhead{m$_{410}$} & 
\colhead{m$_{430}$} & \colhead{\textit{g'}} &  \colhead{m$_{515}$} & \colhead{\textit{r'}} & 
\colhead{m$_{660}$} & \colhead{\textit{i'}} & \colhead{m$_{861}$} & \colhead{\textit{z'}} \\ }
\startdata
W011a & 17.04$\pm$0.07 & 17.11$\pm$0.08 & 17.31$\pm$0.10 & 16.90$\pm$0.05 & 16.92$\pm$0.05 & 16.15$\pm$0.01 & 15.95$\pm$0.02 & 15.29$\pm$0.01 & 15.05$\pm$0.02 & 14.79$\pm$0.00 & 
14.58$\pm$0.01 & 14.57$\pm$0.01 \\
W096 & \nodata & \nodata & \nodata & \nodata & \nodata & 20.24$\pm$0.34 & 20.33$\pm$0.89 & 19.03$\pm$0.18 & 18.31$\pm$0.43 & 17.56$\pm$0.11 & 16.97$\pm$0.18 & \nodata \\
W076 & 20.52$\pm$0.52 & 20.47$\pm$0.66 & 19.81$\pm$0.39 & 20.20$\pm$0.36 & 19.65$\pm$0.22 & 18.84$\pm$0.03 & 18.83$\pm$0.09 & 17.58$\pm$0.01 & 17.25$\pm$0.03 & 16.84$\pm$0.01 & 16.59$\pm$0.02 & 16.59$\pm$0.01 \\
W021 & 18.93$\pm$0.18 & 19.79$\pm$0.51 & 19.66$\pm$0.48 & 18.88$\pm$0.15 & 19.07$\pm$0.18 & 18.13$\pm$0.03 & 17.94$\pm$0.06 & 17.03$\pm$0.01 & 16.73$\pm$0.03 & 16.38$\pm$0.01 & 16.10$\pm$0.01 & 16.10$\pm$0.01 \\
W038 & \nodata & \nodata & \nodata & \nodata & \nodata & 19.84$\pm$0.27 & 18.88$\pm$0.45 & 18.17$\pm$0.12 & 17.63$\pm$0.32 & 16.93$\pm$0.08 & 16.47$\pm$0.14 & \nodata \\
W047a & \nodata & 15.18$\pm$0.17 & \nodata & 14.79$\pm$0.07 & 14.24$\pm$0.05 & 13.75$\pm$0.05 & 13.55$\pm$0.04 & 14.95$\pm$0.17 & 14.44$\pm$0.02 & 13.93$\pm$0.02 & 13.67$\pm$0.02 & \nodata \\
W005 & \nodata & 17.76$\pm$0.57 & \nodata & 17.16$\pm$0.22 & 16.22$\pm$0.13 & 15.40$\pm$0.11 & 14.98$\pm$0.07 & 17.27$\pm$0.49 & 16.45$\pm$0.06 & 15.61$\pm$0.04 & 15.17$\pm$0.04 & \nodata \\
W016 & \nodata & 14.21$\pm$0.11 & \nodata & 14.02$\pm$0.05 & 13.62$\pm$0.04 & 13.35$\pm$0.04 & 13.20$\pm$0.03 & 14.12$\pm$0.12 & 13.78$\pm$0.02 & 13.44$\pm$0.01 & 13.26$\pm$0.01 & \nodata \\
\enddata
\tablecomments{
The error bars denoted by 0.00 represent magnitude errors lower than 0.01 mag. The {\it T80S} sources are the same ones appearing in Table 2.
}
\end{deluxetable*}
\end{longrotatetable}

\begin{longrotatetable}
\begin{deluxetable*}{lcccccccccccc}
\label{tab:lista7}
\tablecaption{Photometric data for {\it T80S} sources that are suggested to be new members of CMa R1, identified in the present work (see Table \ref{tab:lista4})}
\tablewidth{700pt}
\tabletypesize{\scriptsize}
\tablehead{
\colhead{Gaia-DR2} &\colhead{\textit{u'}} & \colhead{m$_{378}$} &  \colhead{m$_{395}$} & \colhead{m$_{410}$} & 
\colhead{m$_{430}$} & \colhead{\textit{g'}} &  \colhead{m$_{515}$} & \colhead{\textit{r'}} & 
\colhead{m$_{660}$} & \colhead{\textit{i'}} & \colhead{m$_{861}$} & \colhead{\textit{z'}} \\
}
\startdata
3046062725556787200 & 17.62$\pm$0.07 & 17.63$\pm$0.07 & 17.88$\pm$0.10 & 17.30$\pm$0.05 & 17.07$\pm$0.03 & 16.31$\pm$0.01 & 16.06$\pm$0.01 & 15.31$\pm$0.00 & 15.08$\pm$0.01 & 14.81$\pm$0.00 & 14.60$\pm$0.00 & 14.60$\pm$0.00 \\
3046107977332044032& 14.36$\pm$0.01 & 14.54$\pm$0.01 & 14.86$\pm$0.01 & 14.49$\pm$0.01 & 14.46$\pm$0.01 & 14.07$\pm$0.00 & 13.87$\pm$0.00 & 13.57$\pm$0.00 & 13.44$\pm$0.00 & 13.36$\pm$0.00 & 13.29$\pm$0.00 & 13.34$\pm$0.00 \\
3046013792988609408 & 21.49$\pm$1.12 & 20.58$\pm$0.68 & 20.23$\pm$0.53 & 22.04$\pm$1.66 & 21.12$\pm$0.76 & 19.91$\pm$0.08 & 19.77$\pm$0.19 & 18.50$\pm$0.02 & 18.05$\pm$0.05 & 17.45$\pm$0.01 & 17.03$\pm$0.02 & 17.00$\pm$0.02 \\
3046064958939721728 & 17.60$\pm$0.08 & 17.51$\pm$0.09 & 18.03$\pm$0.15 & 17.49$\pm$0.07 & 17.41$\pm$0.06 & 16.77$\pm$0.01 & 16.56$\pm$0.02 & 15.97$\pm$0.01 & 15.75$\pm$0.02 & 15.54$\pm$0.01 & 15.36$\pm$0.01 & 15.35$\pm$0.01 \\
304606330967231296 & 19.66$\pm$0.28 & 18.98$\pm$0.19 & 19.16$\pm$0.24 & 20.33$\pm$0.48 & 19.23$\pm$0.17 & 18.44$\pm$0.03 & 18.26$\pm$0.06 & 17.15$\pm$0.01 & 16.83$\pm$0.02 & 16.44$\pm$0.01 & 16.11$\pm$0.01 & 16.10$\pm$0.01 \\
. & . & . & . & . & . & . & . & . & . & . & . & . \\
. & . & . & . & . & . & . & . & . & . & . & . & . \\
. & . & . & . & . & . & . & . & . & . & . & . & . \\
\enddata
\tablecomments{
The error bars denoted by 0.00 represent magnitude errors lower than 0.01 mag. For illustration, only a part of the table is presented. The entire list of objects is available on-line.
}
\end{deluxetable*}
\end{longrotatetable}

\begin{longrotatetable}
\begin{deluxetable*}{lcccccccccccc}
\label{tab:lista8}
\tablecaption{Photometric data for {\it T80S} sources  identified in the present work as candidates that we suggested  to be possible members of CMa R1  (see Table \ref{tab:lista5})}
\tablewidth{700pt}
\tabletypesize{\scriptsize}
\tablehead{
\colhead{Gaia-DR2} &\colhead{\textit{u'}} & \colhead{m$_{378}$} &  \colhead{m$_{395}$} & \colhead{m$_{410}$} & 
\colhead{m$_{430}$} & \colhead{\textit{g'}} &  \colhead{m$_{515}$} & \colhead{\textit{r'}} & 
\colhead{m$_{660}$} & \colhead{\textit{i'}} & \colhead{m$_{861}$} & \colhead{\textit{z'}} \\
}
\startdata
304606736412144384 & 16.12$\pm$0.02 & 16.13$\pm$0.02 & 16.32$\pm$0.03 & 16.07$\pm$0.02 & 16.00$\pm$0.01 & 15.56$\pm$0.00 & 15.32$\pm$0.01 & 14.85$\pm$0.00 & 14.69$\pm$0.01 & 14.46$\pm$0.00 & 14.28$\pm$0.00 & 14.28$\pm$0.00 \\
3046092820387743360 & 22.07$\pm$2.25 & 22.84$\pm$9.61 & 20.40$\pm$1.07 & \nodata & 19.78$\pm$0.39 & 20.57$\pm$0.26 & 19.61$\pm$0.29 & 19.19$\pm$0.09 & 18.82$\pm$0.25 & 17.96$\pm$0.04 & 17.45$\pm$0.05 & 17.37$\pm$0.04 \\
3046067879517494272& 18.47$\pm$0.10 & 18.42$\pm$0.11 & 18.44$\pm$0.12 & 18.47$\pm$0.09 & 18.51$\pm$0.09 & 17.90$\pm$0.02 & 17.59$\pm$0.03 & 17.04$\pm$0.01 & 16.84$\pm$0.02 & 16.54$\pm$0.01 & 16.31$\pm$0.01 & 16.30$\pm$0.01 \\
3046092515449873024& 15.99$\pm$0.02 & 15.92$\pm$0.02 & 15.99$\pm$0.02 & 15.81$\pm$0.02 & 15.76$\pm$0.01 & 15.40$\pm$0.00 & 15.16$\pm$0.01 & 14.82$\pm$0.00 & 14.69$\pm$0.01 & 14.49$\pm$0.00 & 14.34$\pm$0.00 & 14.35$\pm$0.00 \\
3046091759535625856 & 12.52$\pm$0.00 & 12.43$\pm$0.00 & 12.47$\pm$0.00 & 12.51$\pm$0.00 & 12.52$\pm$0.00 & 12.35$\pm$0.00 & 12.19$\pm$0.00 & 12.00$\pm$0.00 & 11.92$\pm$0.00 & 11.82$\pm$0.00 & 11.72$\pm$0.00 & 11.72$\pm$0.00 \\
. & . & . & . & . & . & . & . & . & . & . & . & . \\
. & . & . & . & . & . & . & . & . & . & . & . & . \\
. & . & . & . & . & . & . & . & . & . & . & . & . \\
\enddata
\tablecomments{
The error bars denoted by 0.00 represent magnitude errors lower than 0.01 mag. For illustration, only a part of the table is presented. The entire list of objects is available on-line.
}
\end{deluxetable*}
\end{longrotatetable}

\acknowledgments

This work was partially supported by FAPESP Proc. 2014/18100-4 (JGH),  Proc. 2017/18191-8 (FN), Proc. 2017/19458-8 (AH), Proc. 2018/06822-6 (TSS), 
Proc. 2018/21250-9 (HDP), Proc. 2018/05866-0 (VJP).
JGH acknowledges support from CNPq (305590/2016-6).
We thank Stavros Akras, F\'abio R. Herpich, and Alvaro Alvarez-Candal  for their suggestions and comments on the manuscript.

This work has made use of data from the European Space Agency (ESA) mission
{\it Gaia} (\url{https://www.cosmos.esa.int/gaia}), processed by the {\it Gaia}
Data Processing and Analysis Consortium (DPAC,
\url{https://www.cosmos.esa.int/web/gaia/dpac/consortium}). Funding for the DPAC
has been provided by national institutions, in particular the institutions
participating in the {\it Gaia} Multilateral Agreement.

The S-PLUS project, including the T80-South robotic telescope and the S-PLUS scientific survey, was founded as a partnership between the Funda\c{c}\~{a}o de Amparo \`{a} Pesquisa do Estado de S\~{a}o Paulo (FAPESP), the Observat\'{o}rio Nacional (ON), the Federal University of Sergipe (UFS), and the Federal University of Santa Catarina (UFSC), with important financial and practical contributions from other collaborating institutes in Brazil, Chile (Universidad de La Serena), and Spain (Centro de Estudios de F\'{\i}sica del Cosmos de Arag\'{o}n, CEFCA). We further acknowledge financial support from the São Paulo Research Foundation (FAPESP), the Brazilian National Research Council (CNPq), the Coordination for the Improvement of Higher Education Personnel (CAPES), the Carlos Chagas Filho Rio de Janeiro State Research Foundation (FAPERJ), and the Brazilian Innovation Agency (FINEP).

The authors who are members of the S-PLUS collaboration are grateful for the contributions from CTIO staff in helping in the construction, commissioning and maintenance of the T80-South telescope and camera. We are also indebted to Rene Laporte and INPE, as well as Keith Taylor, for their important contributions to the project.
From CEFCA, we thank Antonio Mar\'{i}n-Franch for his invaluable contributions in the early phases of the project, David Crist{\'o}bal-Hornillos and his team for their help with the installation of the data reduction package \textsc{jype} version 0.9.9, C\'{e}sar \'{I}\~{n}iguez for providing 2D measurements of the filter transmissions, and all other staff members for their support with various aspects of the project. 






\end{document}